\documentclass[a4paper,11pt]{article}
\usepackage{jinstpub} 

\usepackage{subcaption}
\usepackage{booktabs}
\usepackage{appendix} 

\usepackage{cleveref}

\usepackage{lineno}

\title{\boldmath Active Edge Silicon Sensors Fabricated With Edge Ion Implantation and Microwave Annealing for Dopant Activation}

\author[a,1]{A. Gentry, \note{Corresponding author.}}
\author[b]{J. Segal,}
\author[b]{A. Kok,}
\author[b]{C. Kenney,}
\author[a]{S. Seidel}
\affiliation[a]{Department of Physics and Astronomy, University of New Mexico,\\ 210 Yale Blvd.~NE, Albuquerque, NM 87106, U.S.A.}
\affiliation[b]{SLAC National Accelerator Laboratory, 2575 Sand Hill Rd, Menlo Park, CA 94025}

\emailAdd{agentry2@unm.edu}

\abstract{Silicon detectors typically require an insensitive area around their periphery to accommodate guard rings, which help maintain the electric field uniformity around edge pixels and isolate the high leakage current from the physical edges of the detector. Minimization of this insensitive region is desirable for applications in high-energy physics, X-ray experiments, and medical imaging. Existing active edge technology offers a solution for reduction or total elimination of the insensitive region, via a continuation of the highly doped backside up the sidewalls of the device. However, current methods for realizing this technology are complex and expensive. We propose a new technique that simplifies the fabrication of highly doped edges using side ion implantation and microwave annealing. Tests demonstrating the feasibility of this proposed process were performed on a set of sensors, and current versus bias voltage measurements probing the edge effects were performed before and after the edge implantation and annealing. To aid in interpretation of the results, TCAD simulations of the test devices were performed. Significant improvement in the edge leakage current is observed, indicating the promise of this simplified process for fabrication of active edge sensors. }


\arxivnumber{2605.05453} 

\begin{document}
\maketitle
\flushbottom

\section{Introduction}
\label{sec:intro}

Silicon radiation imaging detectors usually require several structures surrounding the sensitive area. Closest to the sensitive region is typically a current collecting guard ring, which helps to improve the uniformity of the electric field near edge pixels and collects current originating from the periphery. This includes potentially large currents originating from the edge of the device, where the saw cut during dicing can cause damage. Outside of the current-collecting ring, there are typically several concentric floating guard rings, which increase the breakdown voltage. Finally, there is space between the outermost guard ring and the cut edge of the device, in order to reduce current from the saw cut. The width of the insensitive region depends on the design parameters of the device, but it typically ranges from a few hundreds of microns to several millimeters \cite{Benoit:2009zz}.

\begin{figure}[htbp]
\centering
        \includegraphics[width=.65\textwidth,trim={0 0 0 0},clip]{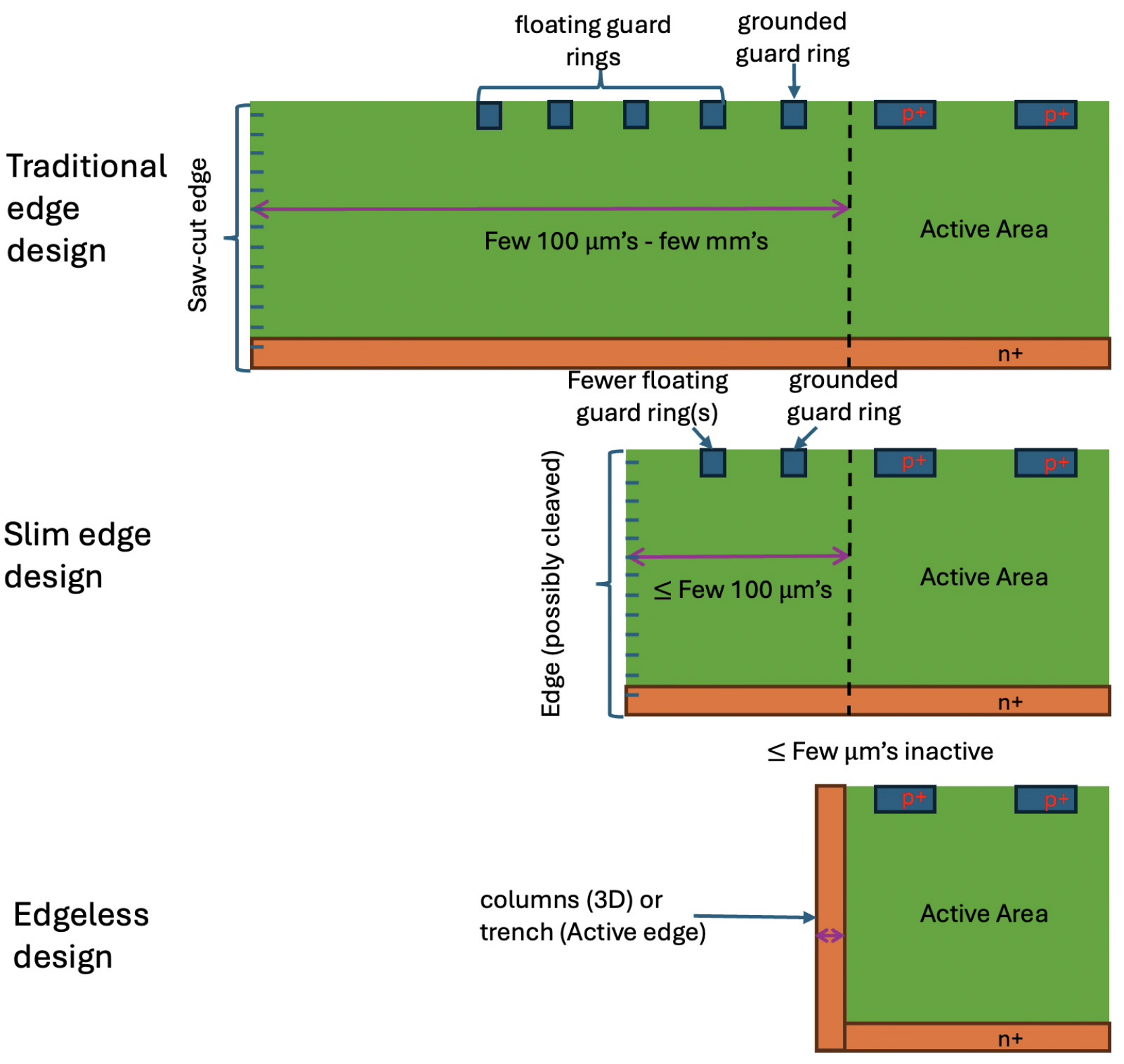}

\caption{Diagrams showing the layout of silicon detectors with traditional edge structure, slim edges, or edgeless layout. The diagrams are not to scale. The right side of each device shows two pixels near the edge of the active area. The left side of each device shows the typical structures  along the device periphery for different edge designs.}
\label{fig:Edges}
\end{figure}

For many applications, it is desirable to reduce or even eliminate this insensitive region, and so there is interest in designing ``slim edge," which require less extra space and/or fewer guard rings, or even ``edgeless" radiation detectors. A diagram demonstrating the difference between traditional edges, slim edges, and edgeless devices can be seen in Figure \ref{fig:Edges}.  In high-energy collider experiments, the increased active-area coverage and reduced material budget of such devices are highly desirable. In X-ray applications, active edges can also enable edge-on detection, increasing effective device thickness and therefore the detectable photon energy range \cite{6154338,Bates:2013ixa}. Several solutions have been studied to achieve such devices, such as the scribe-cleave-passivate (SCP) process \cite{FADEYEV2013260,Fadeyev:2014uaa}, sensors with edge current terminating structures \cite{Ruggiero:2007zzd}, 3D sensors \cite{Parker:1996dx}, and active edges \cite{Kenney20012405,KENNEY2006272}, with varying degrees of success.

In an active edge device, the doped region on the backside extends to the sidewalls. The fabrication of active edge devices typically involves complicated and expensive processing \cite{DallaBetta:2016dqn}. For example, it requires wafer bonding to a support wafer, deep trench etches, diffusive doping, and trench filling. Nonetheless, the active edge process is an effective way of fabricating planar sensors with sensitivity up to the edge of the side implant \cite{Meschini:2016bkd}. Such devices were studied extensively for use at the Large Hadron Collider (LHC), but due to these challenges was ultimately not mass produced for use as a major component of any of the LHC experiments \cite{ERANEN200985,BOMBEN201341,Terzo:2021zov}. 3D sensors, which were developed for radiation hardness and inherently have a slim or no edge, will be used in the upgrades to the tracking detectors of the two largest LHC experiments, ATLAS and CMS, despite the challenges in fabrication, motivated primarily by the need for extreme radiation hardness in the inner layers \cite{ATLAS:2017svb,CMS:2012sda}. Alternative approaches which attempt to simplify the process of making active edge devices have been tested. One example is the ``perforated edge" approach \cite{KOYBASI2020163176}, which precludes the need for a support wafer by etching many separated trenches rather than one trench around the entire device. Another approach used a four sided implant \cite{ERANEN200985}, after attaching a support wafer and etching wide trenches. This approach required the use of non-planar lithography to subsequently fabricate the topside structures.


In this work, we introduce a simplified method for edge doping, in which ion implantation is performed along the device edges after wafer dicing and completion of all prior fabrication steps, including metallization. However, it is necessary to activate the dopant after ion implantation with an annealing step. With standard thermal annealing, this is not possible, as all other fabrication steps must be completed prior to dicing, including metallization. The high temperatures required to activate the dopant would damage the existing structures. 

Microwave annealing (MWA) has been shown to be a viable option for low-temperature activation of dopants and is possible to use even on devices with metal already deposited \cite{Lee2009,Lu2010,Segal:2018cyk,Segal:2021,Tsai2022}. Thus, MWA could enable the fabrication of active edge devices using a 4-sided edge ion implantation.

A feasibility study for this concept was carried out using a set of pixel detectors and diodes. TCAD simulations of these devices were also developed to better understand the effects at the edge of the device and the results of the I-V measurements, both before and after the active edge implantation and anneal.

\section{Test Devices}

\begin{figure}[htbp]
\centering
\begin{subfigure}{0.49\textwidth}
        \includegraphics[width=1\textwidth,trim={18cm 0 0 0},clip]{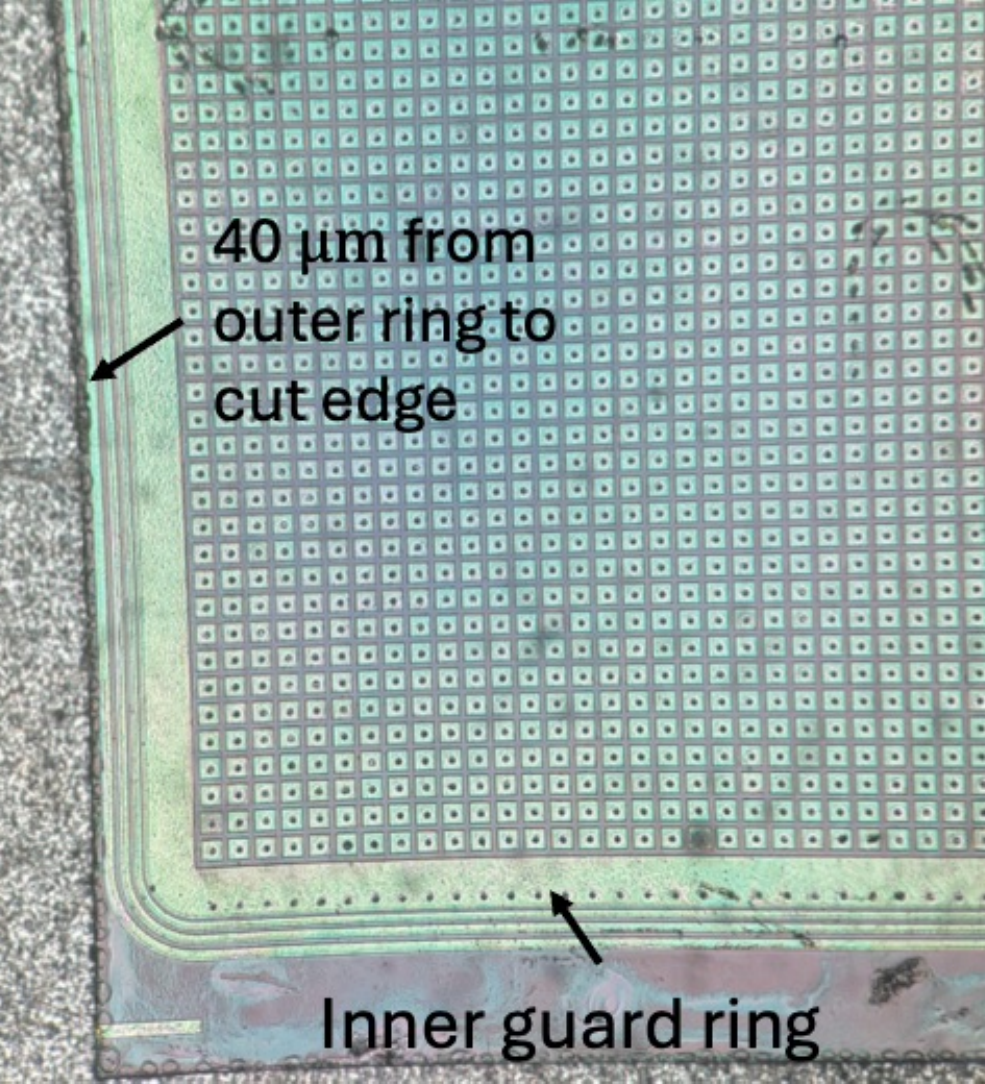}
	\caption{}
	\label{fig:SensorPix}
\end{subfigure}
\begin{subfigure}{.49\textwidth}
        \includegraphics[width=1\textwidth,trim={0 0 18cm 0},clip]{AnnotatedSensors.pdf}
	\caption{}
	\label{fig:SensorRing}
\end{subfigure}
\caption{Photographs showing (a) an example pixel device, and (b) an example ring diode device. Both devices have one edge cut to a width of 40 $\mu m$, as shown in the annotations. For the pixel devices, a single edge pixel and the inner guard ring were simultaneously probed during the I-V measurements. For the ring diodes, the inner circle and innermost ring were simultaneously probed.}
\label{fig:SensorPics}
\end{figure}

The primary goal was to study the efficacy of edge implant and microwave anneal to reduce the effect of high carrier generation due to defects on the cut edge of the device. The devices tested were developed at SLAC as part of several campaigns to build ePix X-ray cameras \cite{Epix2014,Epix2016}. Prototype sized pixel detectors were used, of both p-on-n and n-on-p device polarities. The sensors had array areas $5~\mathrm{mm} \times 5~\mathrm{mm}$ and pixel pitch of either $100~\mathrm{\mu m}~\times 100 ~\mathrm{\mu m}$, $110~\mathrm{\mu m}~\times 110 ~\mathrm{\mu m}$, or $300~\mathrm{\mu m}~\times 300 ~\mathrm{\mu m}$. The full list of devices and the design parameters are in Table \ref{tab:AllDevices}. The devices have either 4 or 5 guard rings, with a metal contact present on the innermost ring. The outer rings are floating. An example of one pixel detector can be seen in Figure \ref{fig:SensorPix}.

\begin{table}[ht]
\centering
\caption{Device parameters for all devices tested.}
\label{tab:AllDevices}
\begin{tabular}{ccccc}
\toprule
Device Number & Polarity & Pitch ($\mathrm{\mu m}^2$) & Thickness ($\mathrm{\mu m}$) & Cut Distance ($\mathrm{\mu m}$) \\
\midrule
1  & p-on-n  & $110 \times 110$  & 350  & 40  \\
2  & p-on-n  & $100 \times 100$  & 350  & 40  \\
3  & p-on-n  & $300 \times 300$  & 350  & 40  \\
4  & p-on-n  & $110 \times 110$  & 300  & 140  \\
5  & p-on-n  & $110 \times 110$  & 300  & 140  \\
6  & p-on-n  & $100 \times 100$  & 300  & 140  \\
7  & p-on-n  & $300 \times 300$  & 300  & 140  \\
8  & p-on-n  & $110 \times 110$  & 300  & 240  \\
9  & p-on-n  & $110 \times 110$  & 300  & 240  \\
10  & p-on-n  & $100 \times 100$  & 300  & 240  \\
11  & p-on-n  & $300 \times 300$  & 300  & 240  \\
12 & n-on-p  & $100 \times 100$  & 300  & 40  \\
13 & n-on-p  & $100 \times 100$  & 300  & 140  \\
14 & n-on-p  & $100 \times 100$  & 300  & 140  \\
15 & n-on-p  & $100 \times 100$  & 300  & 240  \\
16 & n-on-p  & $100 \times 100$  & 300  & 240  \\
17 & n-on-p  & ring diode  & 300  & 40  \\
18 & n-on-p  & ring diode  & 300  & 140  \\
19 & n-on-p  & ring diode  & 300  & 140  \\
20 & n-on-p  & ring diode  & 300  & 240  \\
21 & n-on-p  & ring diode  & 300  & 240  \\
\bottomrule
\end{tabular}
\end{table}

Several n-on-p 300 $\mathrm{\mu m}$ thick circular diodes with concentric rings were also studied. An image of an example circular diode can be seen in Figure \ref{fig:SensorRing}. 

We utilized existing sensors with traditional edges that were not designed for this experiment, and created new sawcuts at various distances from the outer guard  ring to study the effect of current generated at the cut surface.

\section{Procedure}
\label{sec:Procedure}
To characterize the edge current effects, measurements of leakage current versus bias voltage (I-V) were taken using a Keithley 4200-SCS Parameter Analyzer and biasing the devices from the backside. For the pixel detectors, the current at the inner guard ring is measured. The primary goal of this measurement was to assess the current generated at the saw-cut edge, and the bulk current is an unavoidable background in this measurement. For the circular diodes, the current at the circular diode and at the innermost ring were measured simultaneously. The circular diode current allows for a measurement of the bulk, or area-based, current density, since the circular diode region covers nearly the entire surface inside the innermost ring.

The I-V measurements were taken before and after several fabrication steps. First, the devices were diced along one edge, leaving either 40, 140, or 240 $\mathrm{\mu m}$ from the edge to the outermost guard ring. The other three cut edges had significantly greater than 240 $\mathrm{\mu m}$ of space from the outer ring. Figure \ref{fig:SensorPics} shows two devices cut to 40 $\mathrm{\mu m}$ as an example. Preliminary I-V measurements were then taken of each device after this initial cut.

Next, the edge implantation was performed by Luxience Technologies~\cite{luxience}. The process required the use of 4-inch wafers. The devices were adhered to wafers using a 3 $\mathrm{\mu m}$ thick layer of photoresist spun onto the holder wafer. After attaching the devices, the photoresist was baked to harden it, and the wafer was rinsed in acetone to remove any resist that may cover the side walls of the devices. 

The p-on-n devices received an implantation of phosphorus at 150 keV with a dose of $10^{16}~\mathrm{cm^{-2}}$ at 60$^\circ$ tilt, four times, with a 90$^\circ$ rotation between each iteration, in order to cover all 4 sides of the device. Similarly, the n-on-p devices received a dose of boron of $10^{16}~\mathrm{cm^{-2}}$ at 60$^\circ$ tilt on all four sides, with an energy of 50 keV. 

Next, the devices were removed from the wafers with an acetone bath and microwave annealed using the AXOM Microwave Anneal system made by DSG Technologies \cite{DSG}. The tool utilizes high resistivity silicon susceptor wafers, which are positioned both above and below the device being annealed. The susceptor wafers improve the uniformity of the microwave field and the temperature across the device being annealed. After an $\mathrm{N_2}$ purge step to remove moisture, a microwave power of 3300 W was applied for 90 s, followed by 300 s at 1650 W. This recipe was tuned by DSG, using a high resistivity silicon wafer placed between the susceptor wafers (the same location as the devices during the activation anneal). The test wafer temperature was monitored and the process was tuned such that the test wafer heats up rapidly to a goal temperature of $380 ^\circ$C during the high power stage, and is held at that temperature for 300 s in the low-power stage.

I-V measurements were taken after the implant step but before microwave annealing, as well as after the anneal.

\section{I-V Results}
\label{sec:Results}


\subsection{Results Before Active Edge Fabrication}

\begin{figure}[htbp]
\centering
        \includegraphics[width=.65\textwidth,trim={0 0 0 13cm},clip]{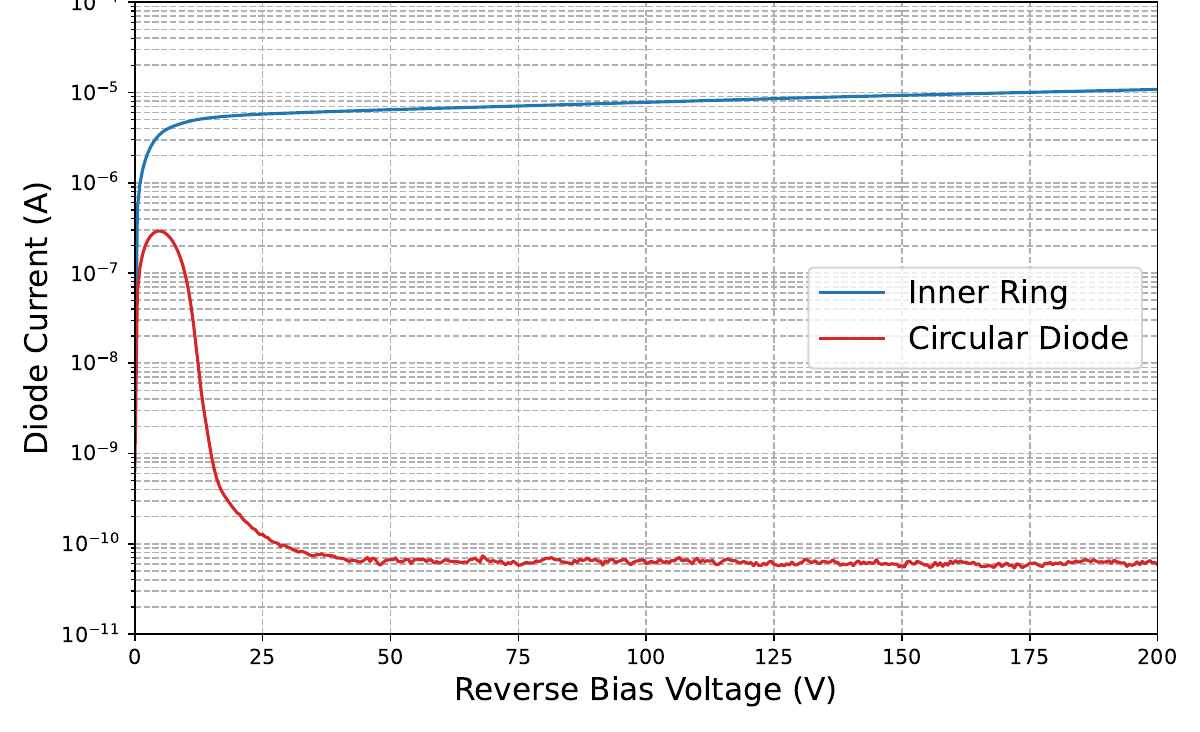}

\caption{Current vs. bias voltage for the innermost ring and circular diode of a circular diode device with edge distance of $140 \mathrm{\mu m}$, prior to edge implantation and MWA.}
\label{fig:CentralDiode}
\end{figure}

I-V measurements on a ring diode are shown in Figure~\ref{fig:CentralDiode}. The circular diode current is larger at low voltage, while the bulk depletes, but is relatively flat after depletion. Since inner ring and circular diode currents are measured simultaneously, the circular diode current should primarily come from the cylindrical region inside of the innermost ring. Thus, the circular diode current allows for a measurement of the bulk current density, which is 3.7 $\mathrm{nA/mm^2}$ in this example. This low bulk current indicates that the ring current is dominated by effects at the saw cut edges. The current collected at the inner ring comprises the edge current, including the guard ring structures and the saw cut. This larger circular diode current at low voltage is most likely due to the fact that prior to full depletion, the inner ring is not fully efficient at removing edge current. After the edge implant and MWA, as will be shown below, the inner ring current is typically reduced significantly.

\begin{figure}[htbp]
\centering
        \includegraphics[width=.65\textwidth,trim={0 0 0 0},clip]{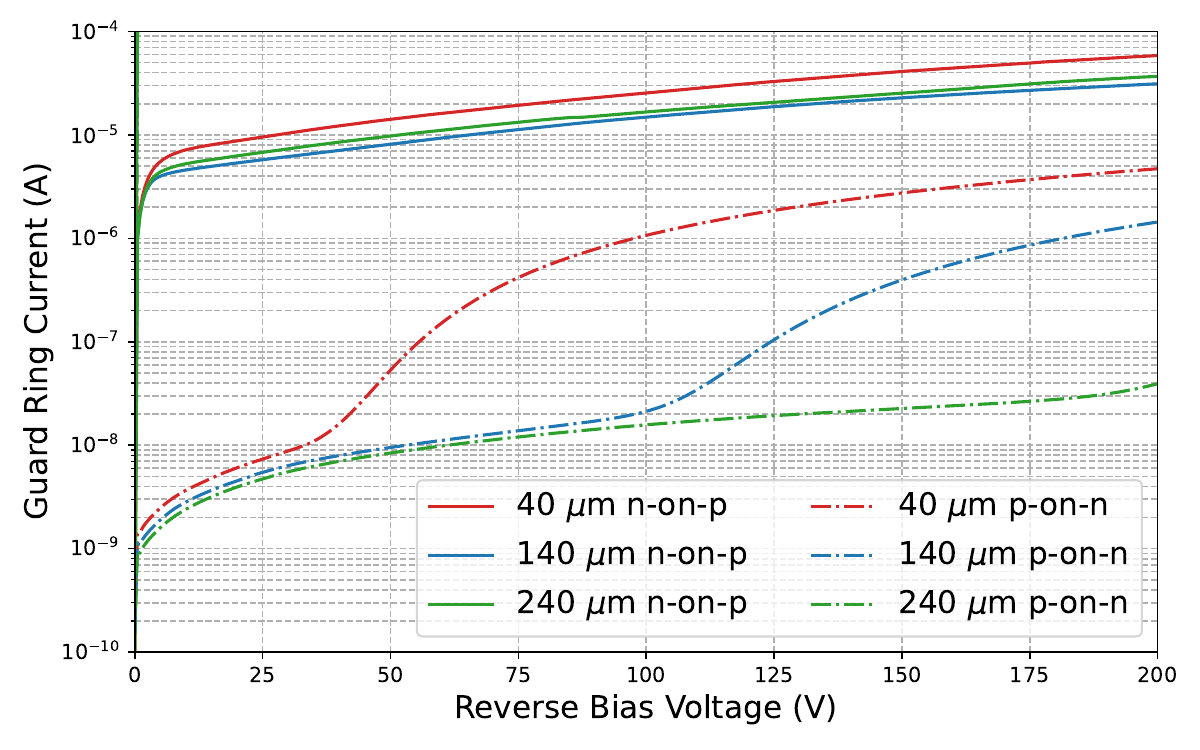}

\caption{Current versus bias voltage plots comparing both n-on-p and p-on-n devices cut at $40, 140, \mathrm{and}~240~\mathrm{\mu m}$, prior to edge implantation and MWA. The n-on-p devices have consistently large currents, regardless of cut distance, while the p-on-n devices exhibit a characteristic current increase depending on cut distance.}
\label{fig:CutCompare}
\end{figure}

The initial pixel sensor guard ring measurements in Figure~\ref{fig:CutCompare} show that the n-on-p devices have large currents already at low voltage, even below the voltage at which the central bulk of the device is expected to reach full depletion. The p-on-n devices, on the other hand, typically have much lower leakage currents, by about 3 orders of magnitude, at lower voltage ($\lesssim 25 ~\mathrm{V})$ for the devices in Figure \ref{fig:CutCompare}. Then, depending on the distance from the outer guard ring to the cut edge, the current increases sharply at some threshold voltage. This effect of current increase with a recognizable difference between cut distances was observed across several different sets of three identical 300~$\mathrm{\mu m}$ or 350~$\mathrm{\mu m}$ thick p-on-n pixel detectors. We propose that this current increase occurs at the voltage at which the edge of the device is depleted. Possible explanations for the differences between p-on-n and n-on-p devices will be explored in Section \ref{sec:TCAD}.

\subsection{Comparison After Active Edge Fabrication}

\subsubsection{P-on-N Devices
}
\begin{figure}[htbp]
\centering
\begin{subfigure}{0.49\textwidth}
        \includegraphics[width=1\textwidth,trim={0 13cm 0 0},clip]{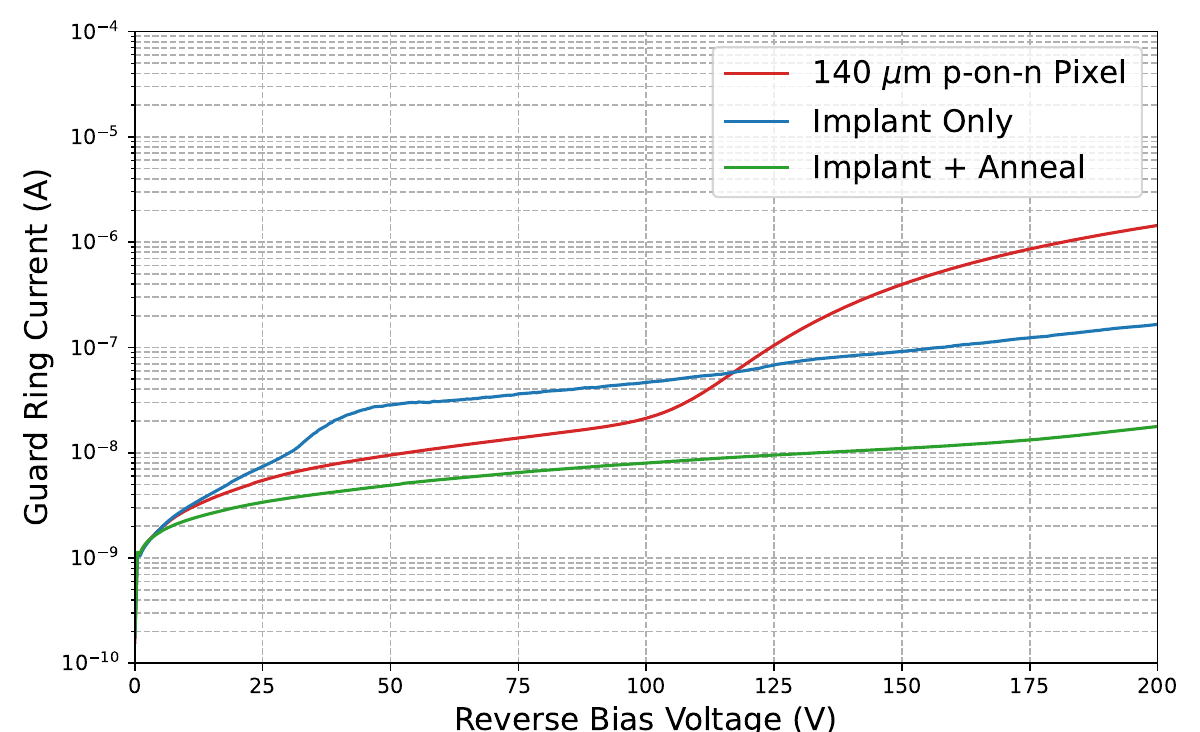}
	\caption{}
    \vspace{.5cm}
	\label{fig:POnNExample}
\end{subfigure}
\hfill
\begin{subfigure}{.49\textwidth}
        \includegraphics[width=1\textwidth,trim={0 0 0 0},clip]{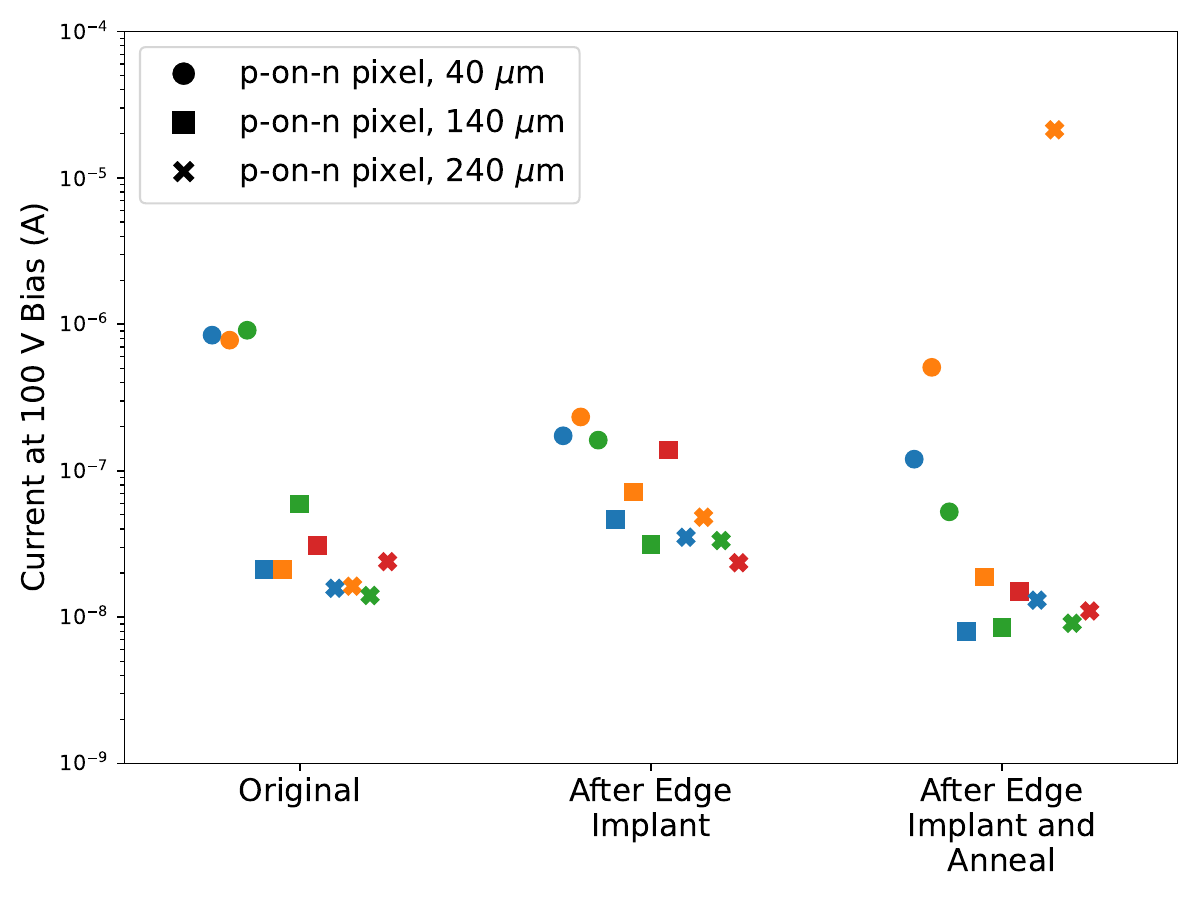}
	\caption{}
	\label{fig:POnNStack}
\end{subfigure}

\caption{(a) Example current versus bias voltage plots for a p-on-n pixel device, with a curve before the edge process, after implantation, and after the MWA activation. (b) Current  at 100 V bias for the 11 p-on-n pixel devices at each of the 3 stages.  Different marker shapes indicate the three edge widths between the saw cut and the outer guard ring. $40~\mathrm{\mu m}$ cut devices exhibited the largest current decrease. } 
\label{fig:ExampleIVs}
\end{figure}

The results of the I-V measurement at each of the 3 stages of the edge process (trimmed edge device, post-implantation, post-MWA) for a p-on-n pixel sensor with 140 $\mathrm{\mu m}$ cut distance can be seen in Figure~\ref{fig:POnNExample}. After implantation and before MWA, the current is larger at low voltages, but does not exhibit the increase associated with the depletion region extending to the edge, indicating that already some dopant is electrically active and buffers the depletion region from reaching the edge. One explanation for the increase in current at low voltage is implantation damage, which is subsequently reduced by the microwave anneal. After the anneal, the current is significantly lower across the full bias range up to 200 V. In the regime where the current has increased (above $\sim$ 150 V in Figure \ref{fig:POnNExample}) the leakage current is lower by nearly 2 orders of magnitude. 

Figure \ref{fig:POnNStack} shows the current at 100 V bias for all 11 p-on-n pixel sensors. The data points are grouped by the 3 different distances between the outer guard ring and the saw cut edge, indicated by different marker shapes. The $40~\mathrm{\mu m}$ devices have significantly larger current at 100 V bias prior to implant and anneal due to the earlier edge depletion, and subsequently exhibited the largest current decrease. There were some outliers, but the general trend shows some decrease in the current after the active edge process. The change is less pronounced in the 140 $\mu \mathrm{m}$ and 240 $\mu \mathrm{m}$ cut devices because 100 V is before the current has significantly increased. The p-on-n devices had a significantly smaller current than the n-on-p devices prior to the edge implantation, and the current remained on average smaller for p-on-n after implantation and MWA.

\subsubsection{N-on-P Devices}
\begin{figure}[htbp]
\centering

\begin{subfigure}{.49\textwidth}
        \includegraphics[width=1\textwidth,trim={0 13cm 0 0},clip]{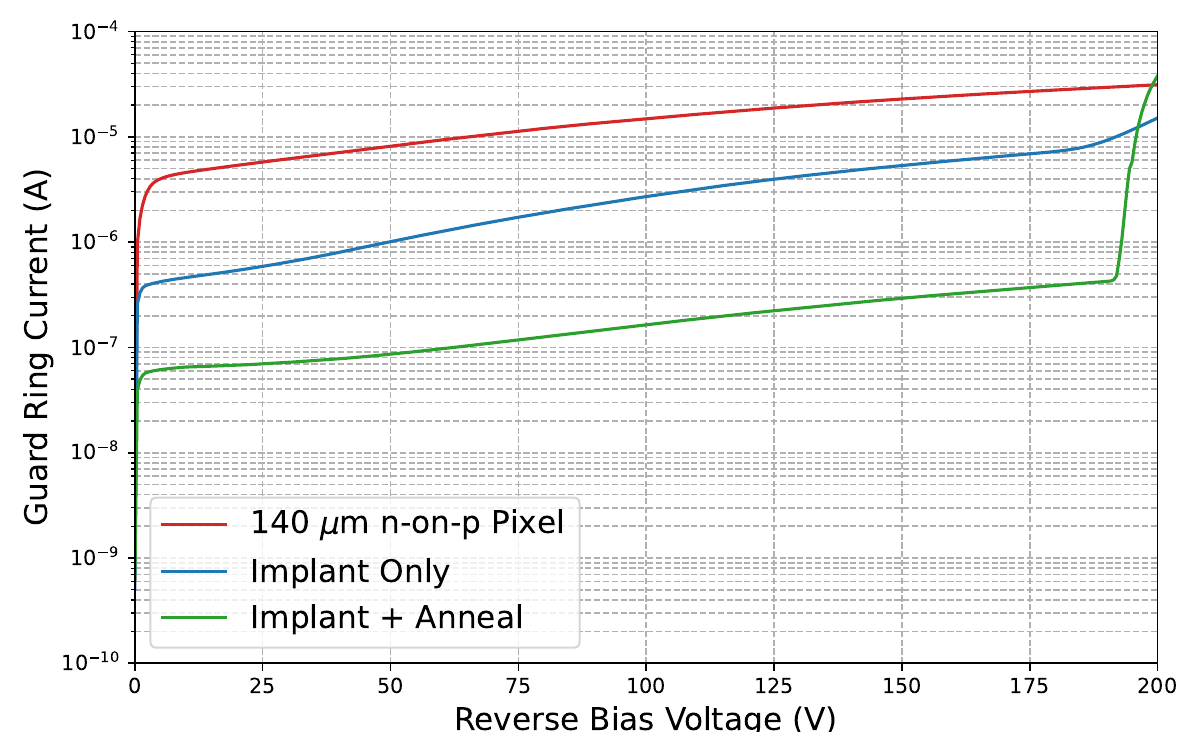}
	\caption{}
	\label{fig:NOnPExample}
\end{subfigure}
\begin{subfigure}{0.49\textwidth}
        \includegraphics[width=1\textwidth,trim={0 13cm 0 0},clip]{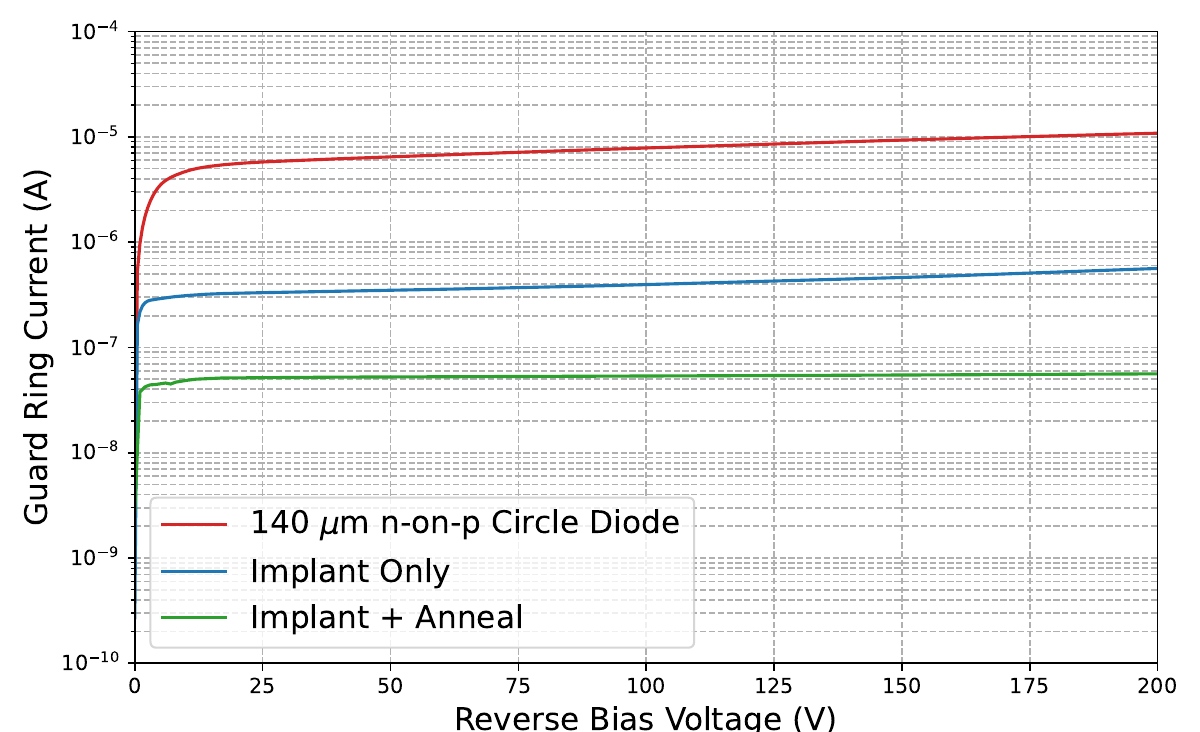}
	\caption{}
	\label{fig:NOnPRingExample}
\end{subfigure}
\begin{subfigure}{0.49\textwidth}
        \includegraphics[width=1\textwidth,trim={0 0 0 0},clip]{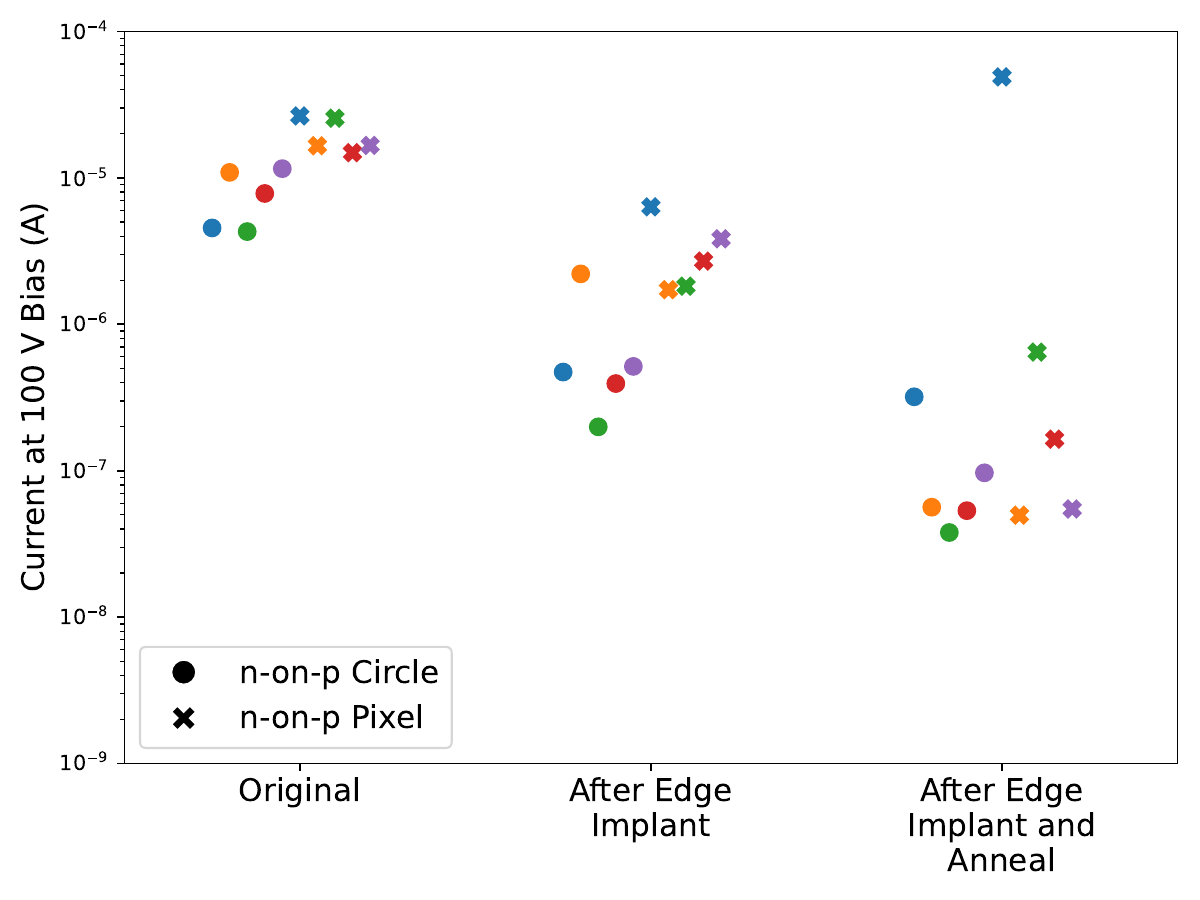}
	\caption{}
	\label{fig:NOnPStack}
\end{subfigure}
\caption{(a) Example leakage current versus bias voltage curves for an n-on-p pixel device, with a curve before the edge process, after implantation, and after the MWA activation. (b) Similar I-V curves for an n-on-p circle diode. (c) Current  at 100 V bias for the 5 n-on-p pixel devices and 5 n-on-p circle diodes at each of the 3 stages. Different marker shapes indicate pixel or circle devices.} 
\label{fig:ExampleIVs}
\end{figure}

 I-V results are shown for an n-on-p pixel sensor in Figure \ref{fig:NOnPExample}. The leakage current decreased by around an order of magnitude after the implantation and another order of magnitude after the anneal. In all cases, throughout most of the voltage range, the current is relatively flat as a function of bias voltage. The device exhibited a sudden increase in the current at about 190 V after the implantation and anneal, returning to nearly the pre-edge implantation current value. The reason for this is not known, but none of the circular diode devices exhibited this early breakdown behavior up to 200 V reverse bias, and there is a significant operational regime between full depletion and 190 V.

Figure \ref{fig:NOnPRingExample} shows an example I-V measurement of an n-on-p ring diode, where the current is from the inner ring, with the circular diode current measured simultaneously but not shown. The results are consistent with those seen for the pixel diodes of the same doping polarity, with typical decrease in guard ring current exceeding two orders of magnitude. 

Figure \ref{fig:NOnPStack} shows the current at 100 V bias for the 5 pixel sensors and 5 circle diodes, indicated by different marker shapes. Since there was no significant difference by edge width, they are not separated by this variable. Most devices exhibited a decrease by about two orders of magnitude after the edge implantation and microwave anneal.

\subsection{MWA Without Implant Results}

\begin{figure}[htbp]
\centering
\begin{subfigure}{0.49\textwidth}
        \includegraphics[width=1\textwidth,trim={0 13cm 0 0},clip]{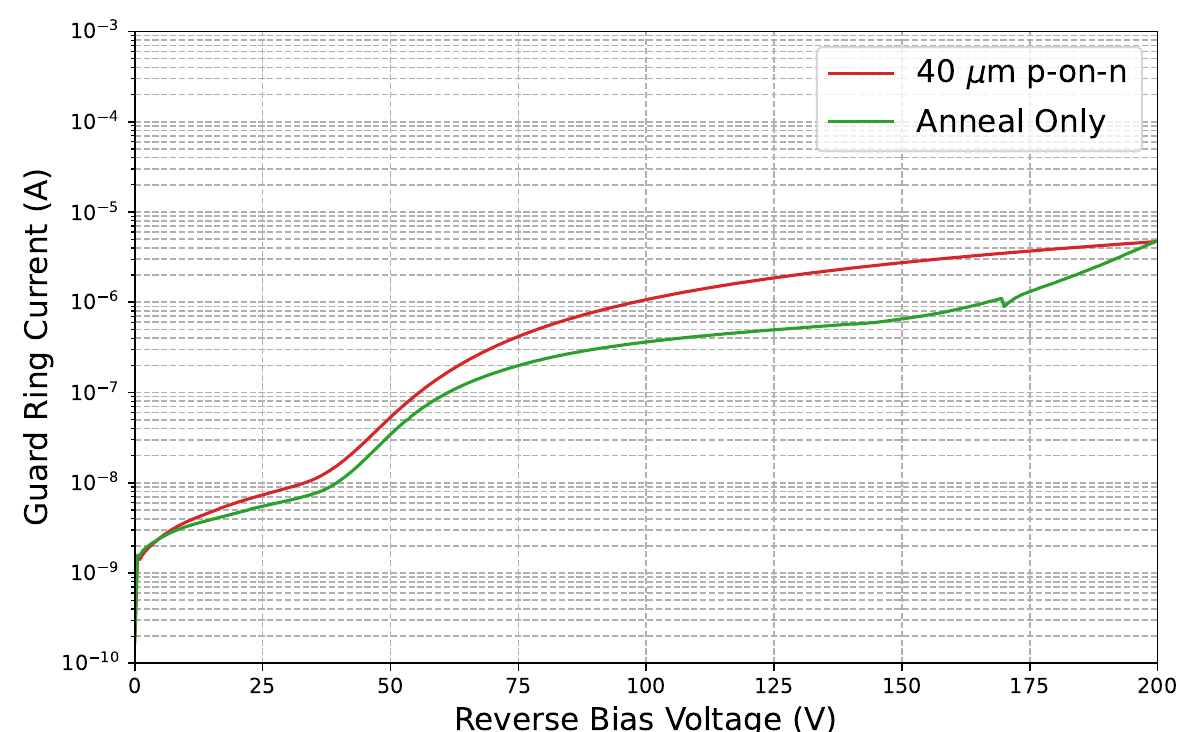}
	\caption{}
	\label{fig:POnNMWaOnly}
\end{subfigure}
\hfill
\begin{subfigure}{.49\textwidth}
        \includegraphics[width=1\textwidth,trim={0 13cm 0 0},clip]{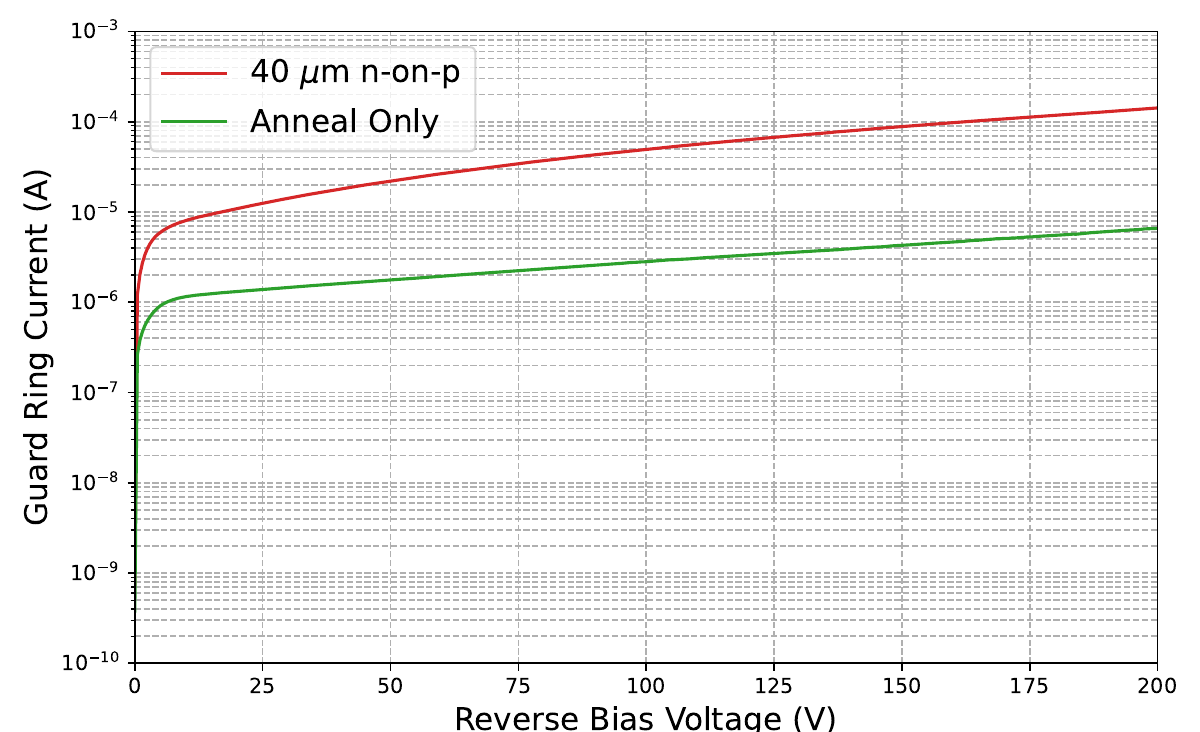}
	\caption{}
	\label{fig:NOnPMWAOnly}
\end{subfigure}
\caption{Current versus bias voltage curves for an (a) p-on-n pixel device, and (b) n-on-p pixel device, which are trimmed to $40~\mathrm{\mu m}$, and then annealed only, without an implant. The result shows a decrease in overall current, however, it is typically significantly less than the decrease observed in the devices where the edge implantation was performed first, indicating that there is still significant current generated from the damaged edge.}
\label{fig:MWAOnlyIVs}
\end{figure}

As discussed in the introduction, our goal is to create an MWA-activated doped region at the sawcut, preventing the depletion of the edge. In addition to this mechanism for current reduction, MWA also has the potential to remove defects; this could also reduce the current. To test for this effect, and assess the relative contribution from the two effects, one device of each polarity was tested with a 40 $\mathrm{\mu m}$ cut and MWA only, without edge implantation. The results can be seen in Figure \ref{fig:MWAOnlyIVs}. The current reduced in both instances, but by less than the total current reduction based on the full implantation + anneal process. The p-on-n device still exhibits a large current jump starting around 40 V reverse bias, indicating that the edge depletion effect is still present without the edge implant to prevent it. This also supports the hypothesis that the current increase is caused by edge depletion.

\section{TCAD Simulations}
\label{sec:TCAD}

To better understand the trends in the detector I-V curves, two-dimensional TCAD simulations were performed using the Sentaurus suite of tools by Synopsis \cite{Sent-Dev}. The device structures were generated in Sentaurus Structure Editor, and a voltage ramp up to 300 V reverse bias was simulated in Sentaurus Device.

These simulations explore the plausibility of the proposed explanation for the qualitative features of the pre-implantation I-V curves shown in Figure \ref{fig:CutCompare}, namely that the current increases when the depletion region expands to the cut edge. In the p-on-n devices, there is a clear current increase occuring at a voltage that increases with increasing edge width, lending support to this proposed mechanism. In the n-on-p devices, it is proposed that the p-spray which was applied to the top surface was insufficient to prevent a channel from forming at the top surface due to positive charges in the oxide, so that the charges generated at the cut edge drift along the top surface even at very low voltage, with little variation in guard ring current due to cut distance. After implantation and MWA activation, enough dopant is activated to buffer the electric field from the edge of the device, so that the poor-quality edge does not contribute significantly to the current up to the measured 200~V reverse bias.

\subsection{Device Structure}
\subsubsection{Topside/bulk}
\label{sec:topside}
\begin{figure}[htbp]
\centering
    \includegraphics[width=.65\textwidth,trim={0 0 0 0},clip]{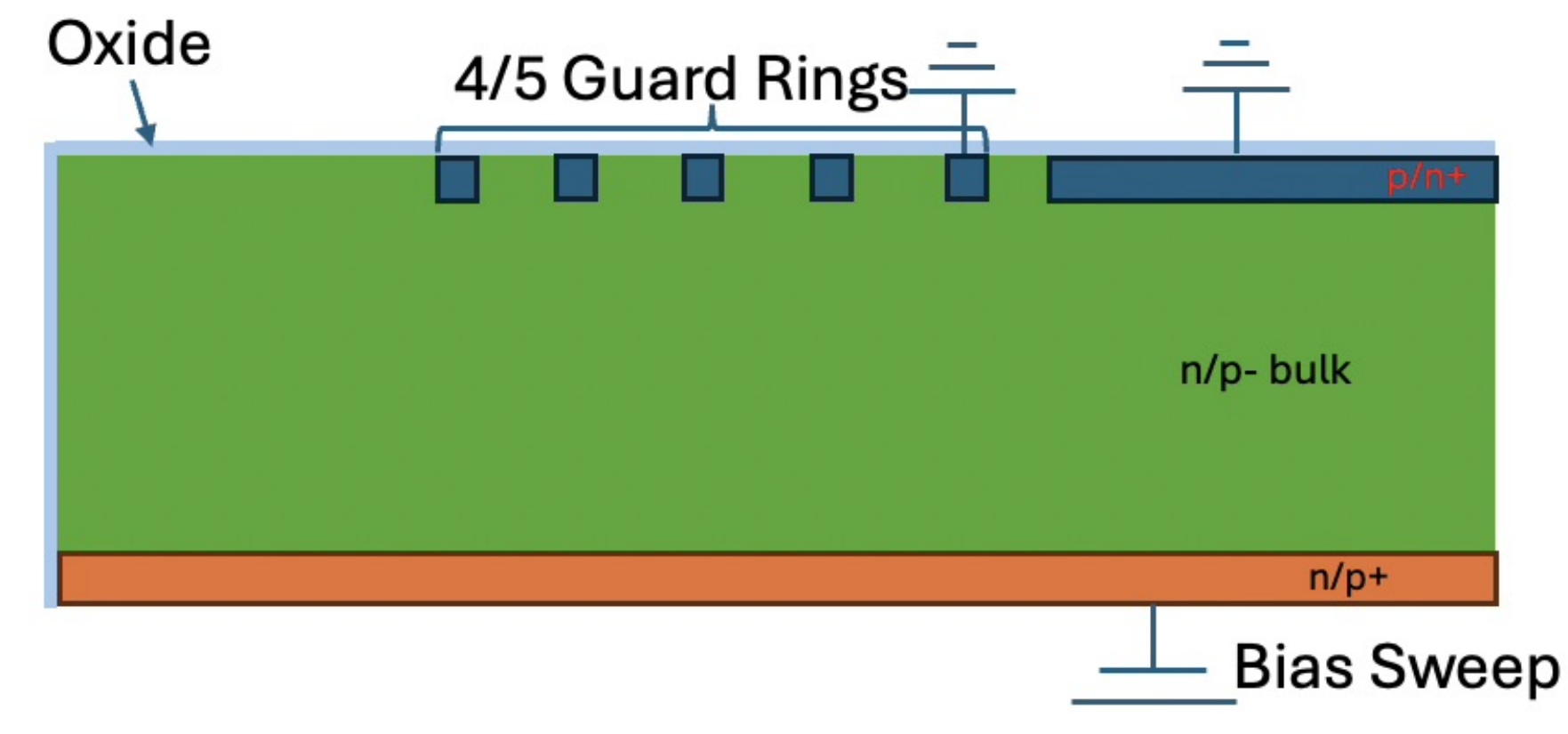}
\caption{A diagram of the simulated devices. The device is not to scale. The distance between the left edge of the leftmost guard ring and the device edge is 40, 140, or 240 $\mathrm{\mu m}$. Not shown, the n-on-p devices also have a p-spray layer on the top-side.}
\label{fig:DeviceDiagram}
\end{figure}

The top- and bottom-side structures were fabricated at SLAC using conventional processing techniques, and so are well known and implemented in the simulation. The simulated devices consisted of a single pad PIN diode with 300 $\mathrm{\mu m}$ thickness, diode width of 500 $\mathrm{\mu m}, $ four (five) concentric guard rings of varying size, with a contact on the innermost guard ring, and the others floating, consistent with the layout of the n-on-p (p-on-n) pixel sensors tested. The guard ring sizes and separation were also chosen to match the pixel devices tested. A diagram (not to scale) of the devices can be seen in Figure \ref{fig:DeviceDiagram}. There is 40, 140, or 240 $\mathrm{\mu m}$ of extra space between the edge of the outer guard ring and the physical edge of the device, to match the dicing of the physical devices which were tested.

 The peak doping concentration for the top and bottom contacts and guard ring regions is $5\times 10^{19} \mathrm{cm^{-3}}$. The doping profile was Gaussian, with its peak at the silicon surface, and width of 0.38 $\mu\mathrm{m}$. The bulk doping concentration in both device polarities is set to $5\times 10^{11}~\mathrm{cm^{-3}}$, which is the quoted bulk doping concentration for the high resistivity float-zone wafers used in the physical sensors. 

The n-on-p devices have a p-spray, which is implemented along the entire top-side of the device. To get a realistic doping profile, a Sentaurus Process simulation limited to the steps relevant to implementing the p-spray was performed, and then an analytical doping profile was defined in the simulated device to match the profile achieved in the Process simulation.

\subsubsection{Sawcut Edge}

The dicing of wafers is known to cause defects along the cut edge \cite{Lee2017M2016366}. Possible explanations for the large current generated at the cut edge are the oxide and interface defects along the edge, or a very rough surface. Further work would be needed to explore the actual mechanism of charge carrier generation at the diced edge. To achieve a large carrier generation at the edge, a 1 $\mu \mathrm{m}$ thick layer of $\mathrm{SiO_2}$ is placed at the left edge of the simulated device, and a high density of mid-bandgap single energy level interface traps is used. Both acceptor-like and donor-like traps were used, with a density of $1\times 10^{13}\mathrm{cm^{-2}}$. A fixed positive interface charge of $1\times 10^{12}~\mathrm{cm^{-2}}$ was also added. The top surface also has a 1 $\mu \mathrm{m}$ thick oxide layer, with defect parameters discussed below.

In preliminary simulations, the density of traps and fixed charge was varied. Higher trap densities generate larger currents. The depletion of the edge can also be affected by the accumulated surface charge in the traps and fixed charge. In p-on-n devices, a net positive charge density will prevent edge depletion, and a net negative charge can cause it to deplete at a lower voltage. However, it is seen that with the configuration of traps used, as long as there are enough acceptor-like traps, their negative charge can cancel out the fixed positive charge, and the donor-like traps have a low occupancy. As long as the fixed oxide charge density is less than the acceptor-like trap density, $N_\mathrm{ox} < N_\mathrm{acc}$, the precise values of $N_\mathrm{ox}, N_\mathrm{don}$ and $N_\mathrm{acc}$ had little effect on the value at which the edge was depleted.

In the n-on-p devices, a net positive charge causes a small depletion region to form along the whole edge, before any bias voltage was applied. Thus, when the depletion region approaches near any part of the edge, the full edge becomes depleted and contributes to the current on the guard ring.

\subsubsection{Edge Doping Profile}
\label{sec:EdgeDoping}
Simulations were performed of devices with and without the edge implanted dopant. Similarly to the process described for the p-spray, the edge doping profile was chosen to match the profile from a Sentaurus Process simulation of the implant applied to the edges of the devices, which is described in Section \ref{sec:Procedure}.

It has been seen, in previous tests with the DSG microwave annealer, that typically a fraction of $< 0.1$ of the dopant is activated, with as low as $\sim 0.002 $ activated in one case~\cite{Segal:2021}, where the activation fraction is the ratio of electrically active dopant to total dopant. It was also seen that minimal dopant diffusion occurs during the microwave annealing. To model the active dopant at the edge, the same doping profile shape is used, with different activation fractions, corresponding to the peak concentration being reduced by a factor from $10^{-5}$ to $10^{-1}$.

\subsection{P-on-N Simulations}

The simulations demonstrate that the qualitative features of the p-on-n profiles are explainable by the depletion region reaching the saw-cut edge, causing the sharp current increase depending on cut distance, and  removal of the edge current contribution after the sidewall is doped. 

\begin{figure}[htbp]
\centering
        \includegraphics[width=.65\textwidth,trim={0 0 0 1.5cm},clip]{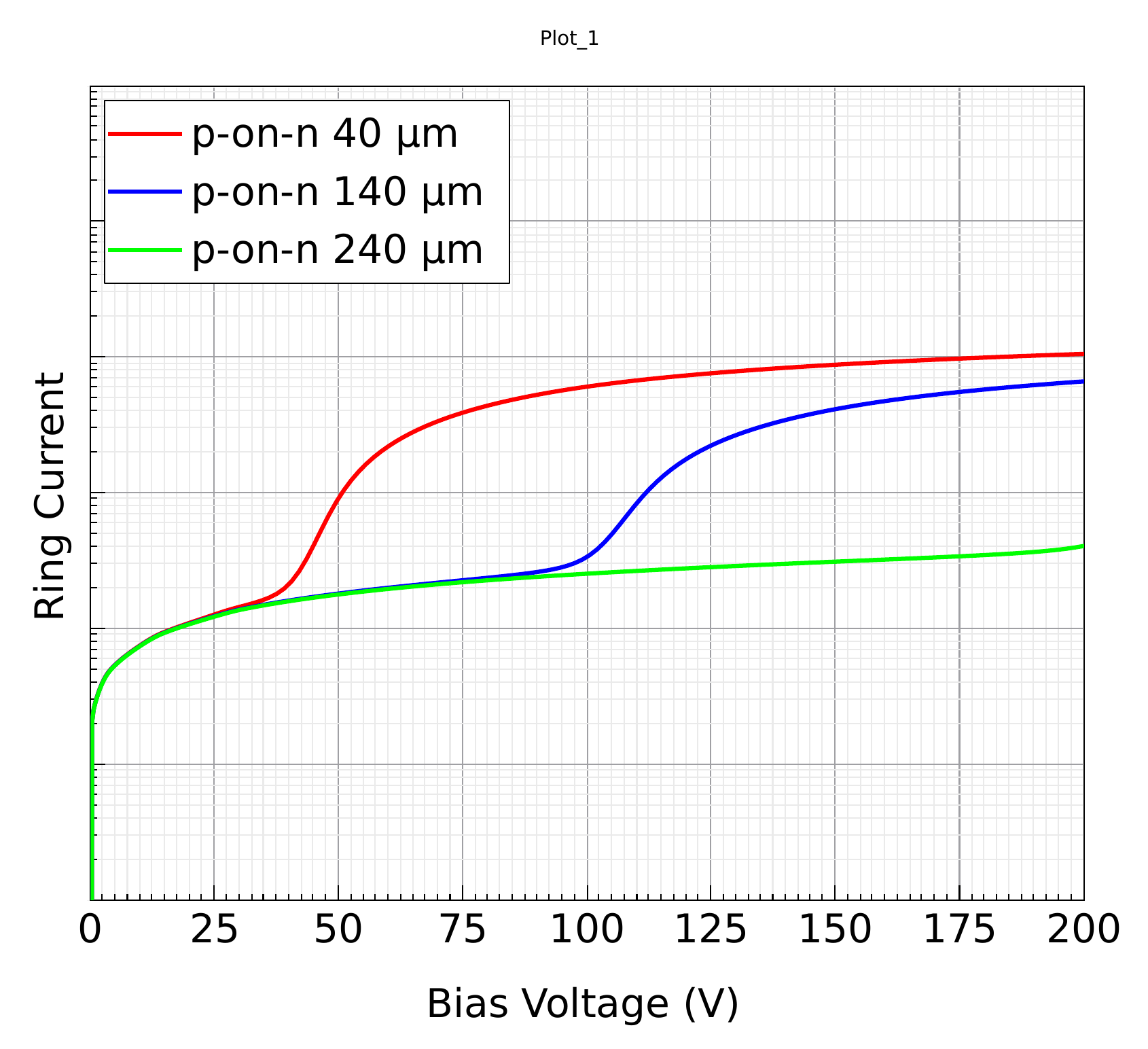}

\caption{Simulated leakage current versus applied bias voltage for p-on-n devices with distance of 40, 140, or 240 $\mathrm{\mu m}$ between the outer guard ring and the edge. They exhibit a low current at low voltage, with a large increase occurring only when the edge is depleted. This simulation matches well the features of the p-on-n sensor data in Figure \ref{fig:CutCompare}.}
\label{fig:PonNIV}
\end{figure}

An I-V plot of the device simulated with the 3 different cut distances is shown in Figure~\ref{fig:PonNIV}. The current exhibits a large increase at a voltage that depends on cut distance. The voltages at which the current increases are similar to those seen in the real data in Figure~\ref{fig:CutCompare}. The scale on the current in Figure \ref{fig:PonNIV} is suppressed. The actual value of the current isn't physically meaningful, because the simulation is 2D and no calibration was done of, for example, the minority carrier lifetime in the bulk in order to have a realistic bulk current density. 

\begin{figure}[htbp]
\centering
\begin{subfigure}{.49\textwidth}
        \includegraphics[width=1\textwidth,trim={0 0 0 0},clip]{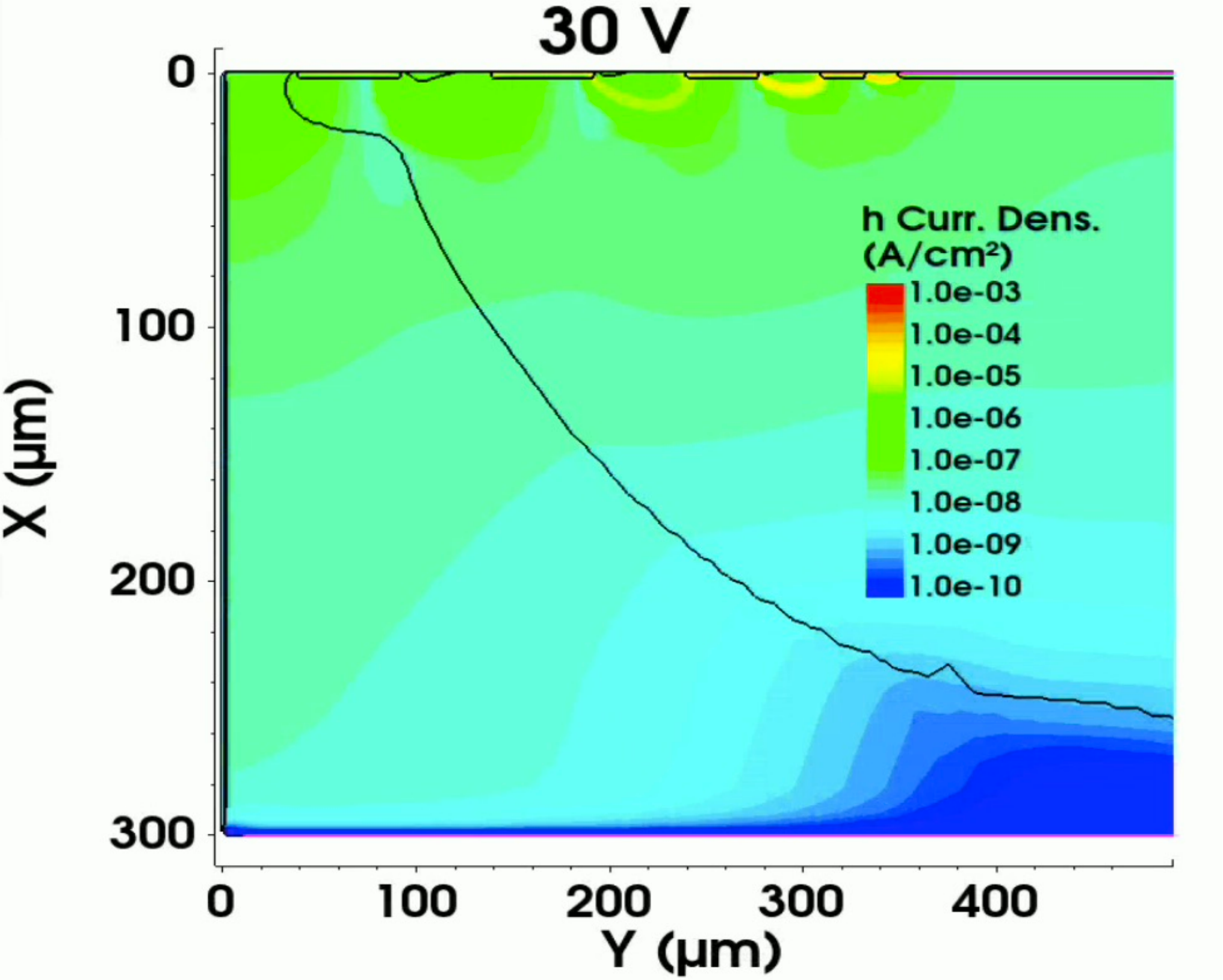}
	\caption{}
	\label{fig:POnNHCurr30V}
\end{subfigure}
\hfill
\begin{subfigure}{.49\textwidth}
        \includegraphics[width=1\textwidth,trim={0 0 0 0},clip]{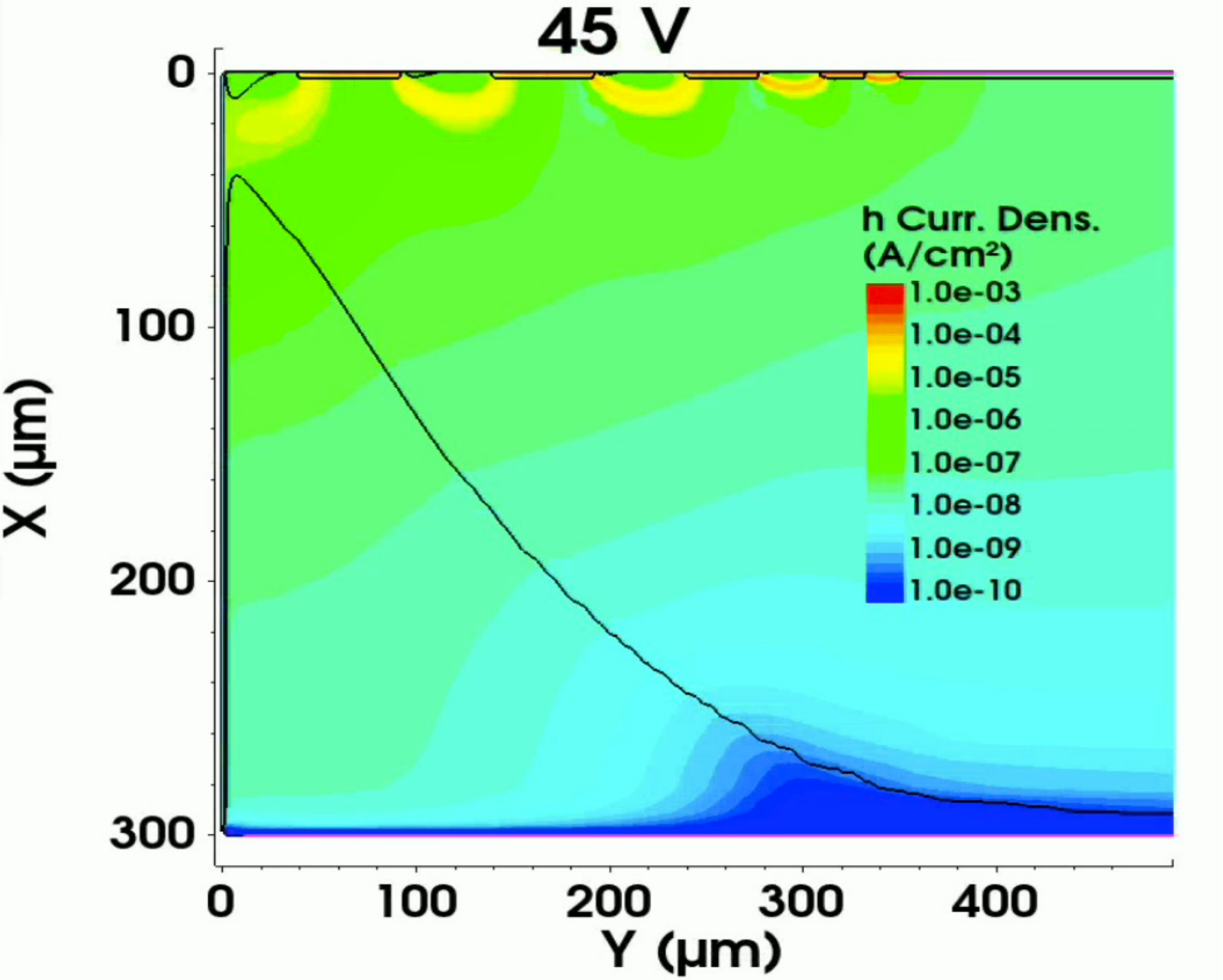}
	\caption{}
	\label{fig:POnNHCurr45V}
\end{subfigure}
\caption{Contour plots of the hole current density in a device with $40~\mathrm{\mu m}$ between the outer guard ring and the device edge, biased at (a) 30 V and (b) 45 V. The black contour lines show the depletion region.}
\label{fig:POnNHCurr}
\end{figure}

To show that it is indeed the edge depletion that corresponds to the current increase, Figure \ref{fig:POnNHCurr} shows the hole current density of the 40 $\mathrm{\mu m}$ cut distance device at 30 V versus 45 V bias. The full structure illustrated in Figure \ref{fig:DeviceDiagram} is simulated, but only the region of the device between the sidewall and the inner guard ring is shown in Figure \ref{fig:POnNHCurr}. The depletion region is shown using a black contour line. Below the current jump, at 30 V, the edge is not yet depleted, and by 45 V the edge has depleted, and a large hole current is flowing from the edge towards the guard rings.

The value where the current increase occurs depends strongly on the bulk doping concentration, as well as the charge concentration at the top surface because both affect the growth of the depletion region with increasing voltage. In the devices shown, $\mathrm{N_{ox} = N_{don} = N_{acc} = 1\times 10^{10}~cm^{-2}}$. Both the acceptor-like and donor-like traps were single energy traps at mid-bandgap. These densities fall within the range typically seen for an $\mathrm{SiO_2-Si}$ interface.

The actual values of the trap/oxide charge density and energy distribution are not known. The edge depletion voltage changes significantly within the margin of error for the bulk doping concentration. However, these simulations show that within a reasonable range of values for the density of surface defects and bulk doping concentration, it is possible to achieve an I-V curve qualitatively matching the data. Hence, the proposed explanation of edge depletion causing the current jump before the edge implantation and MWA is seen to be plausible.

\begin{figure}[htbp]
\centering
        \includegraphics[width=.65\textwidth,trim={0 0 0 1.5cm},clip]{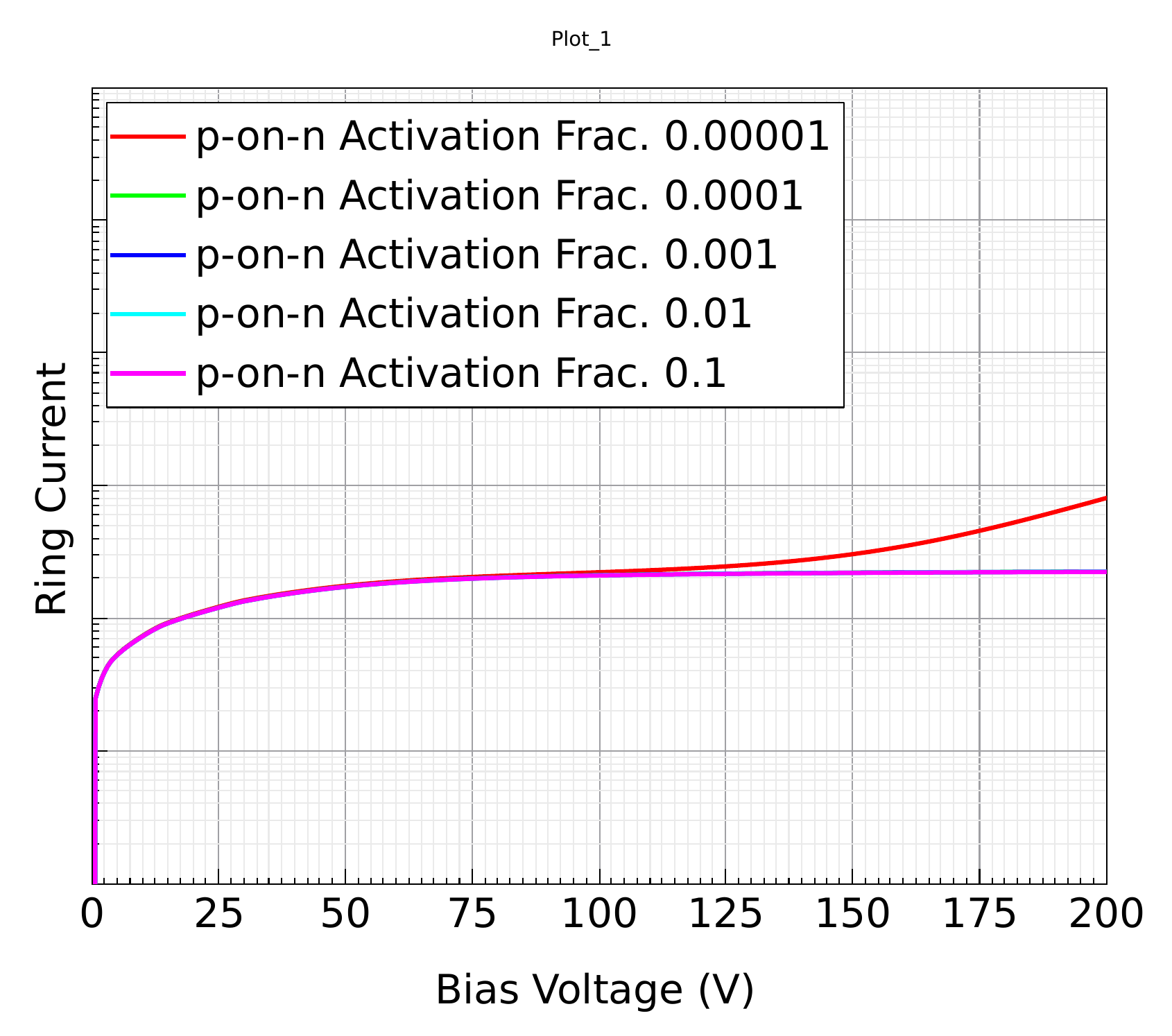}
\caption{Simulated current versus bias voltage curves for $40~\mathrm{\mu m}$ cut distance p-on-n devices with edge implant and varying fraction of the total implanted dopant activated. Only the lowest fraction exhibits edge depletion within the 200 V range. Most of the curves are nearly identical.}
\label{fig:PonNSideImplant}
\end{figure}

We simulated the post-MWA sidewall implant using a range of activated dopant rates, as described in Section \ref{sec:EdgeDoping}. Most of the devices simulated with the n-type sidewall implant exhibited a small leakage current with no current increase, consistent with a total prevention of edge depletion. This is shown in Figure \ref{fig:PonNSideImplant}, which shows the I-V curves for $40~\mathrm{\mu m}$ cut distance devices with activation fractions between $10^{-5}$ and $10^{-1}$. Only the device with activation fraction of $10^{-5}$ exhibits the current increase characteristic of edge depletion.

This result demonstrates that even with the lowest activation fractions previously observed for the MWA activation process, it is reasonable to expect that the contribution to the leakage current from the edge is removed by the edge implant and MWA that is performed.

\subsection{N-on-P Simulations}

Using a surface and edge defect configuration similar to that of the p-on-n simulations, the n-on-p devices exhibited a similar current increase depending on cut distance, corresponding to the edge depletion. This is contrary to what was seen in the data in Figure \ref{fig:CutCompare}. Given the fact that the current decreased by around 2 orders of magnitude after the implantation and anneal, with a significantly smaller change after the MWA only, it is still reasonable to conclude that the large current is related to edge effects. The goal of the n-on-p simulations, then, was to propose a mechanism by which the edge current can contribute at low voltage, with little variation based on cut distance, as seen in Figure \ref{fig:CutCompare}.

A well established problem in n-on-p sensors is electrical shorting along the top surface due to an electron accumulation layer which forms below the positively charged interface. The typical solution to this is a p-spray or p-stop \cite{Batignani:1988as, Goessling:1996fn}. In the n-on-p devices tested, a boron p-spray was applied, with a dose of $4\times 10^{12}~\mathrm{cm^{-2}}$. The  p-spray doping profile was derived from SProcess simulations as described in Section \ref{sec:topside}.

The effect of positive charge present at the top-side $\mathrm{SiO_2-Si}$ interface was explored by varying the defect concentrations in the simulations. It was found that if a positive charge density of about $5\times 10^{11}~\mathrm{cm^{-2}}$ is present at the top-side interface, from a combination of fixed oxide charge and interface traps, a narrow low resistance layer is formed near the surface. The result is that the edge contributes to the current at just a few volts bias. 

To illustrate this effect, one low defect density and one high defect density model are compared, with defect densities defined in Table \ref{tab:defect_models}. The single-energy donor like traps are 0.2 eV above the middle of the bandgap, towards the conduction band, while the acceptor like traps are at mid-bandgap. These defect densities are just above and just below the threshold where the net positive charge is around $5\times 10^{11}~\mathrm{cm^{-2}}$, where some positive charge comes from the fixed charge, $N_\mathrm{ox}$, and some comes from the donor-like traps, which are only partially occupied.

\begin{table}[h]
\centering
\caption{Defect densities for the low and high defect models. Donor-like traps are located 0.2~eV above mid-bandgap (toward the conduction band), while acceptor-like traps are at mid-bandgap.}
\begin{tabular}{lcc}
\hline
 & Low Defect Model & High Defect Model \\
\hline
$N_\mathrm{ox}$ [$\mathrm{cm^{-2}}$] & $2 \times 10^{11}$ & $2 \times 10^{11}$ \\
$N_{\mathrm{acc}}$ [$\mathrm{cm^{-2}}$] & $1 \times 10^{11}$ & $1 \times 10^{11}$ \\
$N_\mathrm{don}$ [$\mathrm{cm^{-2}}$] & $3 \times 10^{11}$ & $5 \times 10^{11}$ \\
\hline
\end{tabular}
\label{tab:defect_models}
\end{table}

\begin{figure}[htbp]
\centering
\begin{subfigure}{.49\textwidth}
        \includegraphics[width=1\textwidth,trim={0 11cm 0 1.5cm},clip]{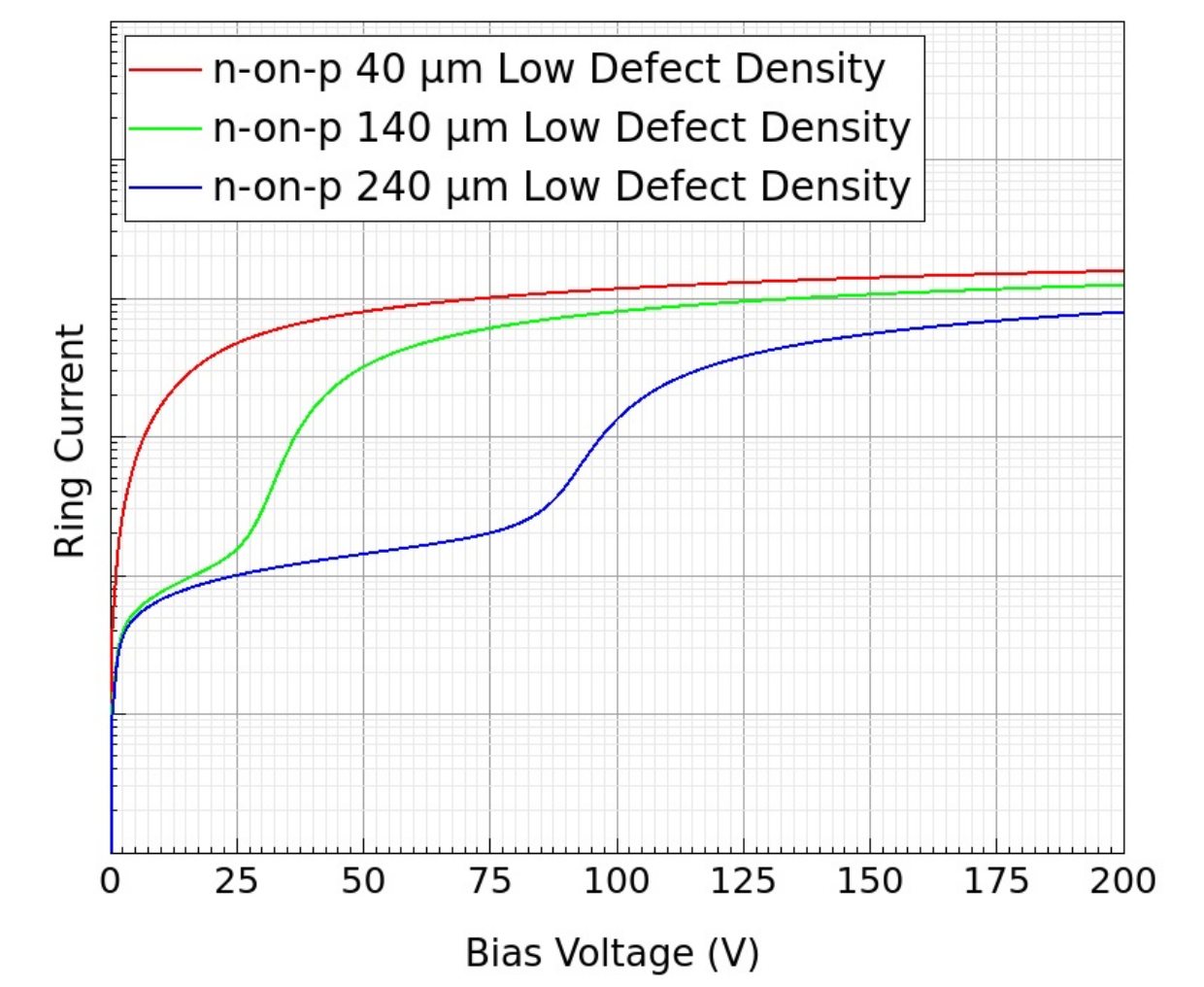}
	\caption{}
	\label{fig:NonPIV1}
\end{subfigure}
\hfill
\begin{subfigure}{.49\textwidth}
        \includegraphics[width=1\textwidth,trim={0 11cm 0 1.5cm},clip]{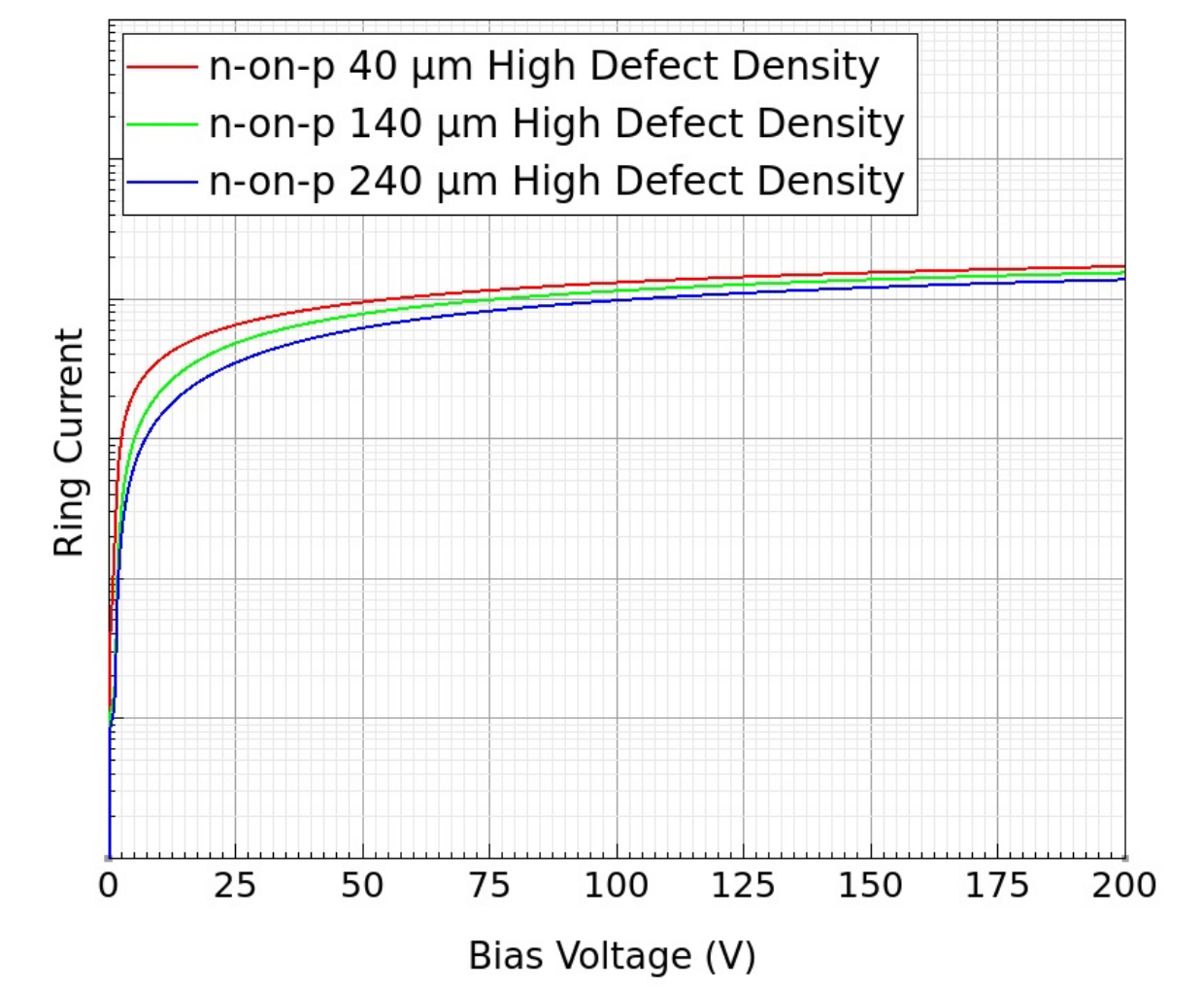}
	\caption{}
	\label{fig:NOnPIV2}
\end{subfigure}
\caption{(a) Simulated current versus bias voltage curves for low defect density n-on-p model, including 3 different cut distances between the guard rings and sensor edge.  (b) Simulated I-V curves for high defect density n-on-p model, including 3 different cut distances between the guard rings and sensor edge. This simulation matches well the features of the n-on-p sensor data in Figure \ref{fig:CutCompare}.}
\label{fig:NonPIV}
\end{figure}

The I-V curves for the low and high defect density can be seen in Figures \ref{fig:NonPIV1} and \ref{fig:NOnPIV2}, respectively. The curves for 40, 140, and 240 $\mathrm{\mu m}$ cut distance are shown for each defect density. In the low defect simulations, the edge depletion depends on the cut distance, as can be seen from the current jump at different voltage depending on the cut distance, comparable to what is seen for p-on-n devices in Figure \ref{fig:PonNIV}. On the other hand, the high defect devices have higher current due to edge depletion at very low voltage, regardless of cut distance. This is because the positive charge concentration at the top surface allows for the charge from the cut edge to be conducted along the surface. The result for the high defect device in Figure \ref{fig:NOnPIV2} matches qualitatively with the data in Figure \ref{fig:CutCompare}.

\begin{figure}[htbp]
\centering
\begin{subfigure}{.49\textwidth}
        \includegraphics[width=1\textwidth,trim={0 0 0 0.1cm},clip]{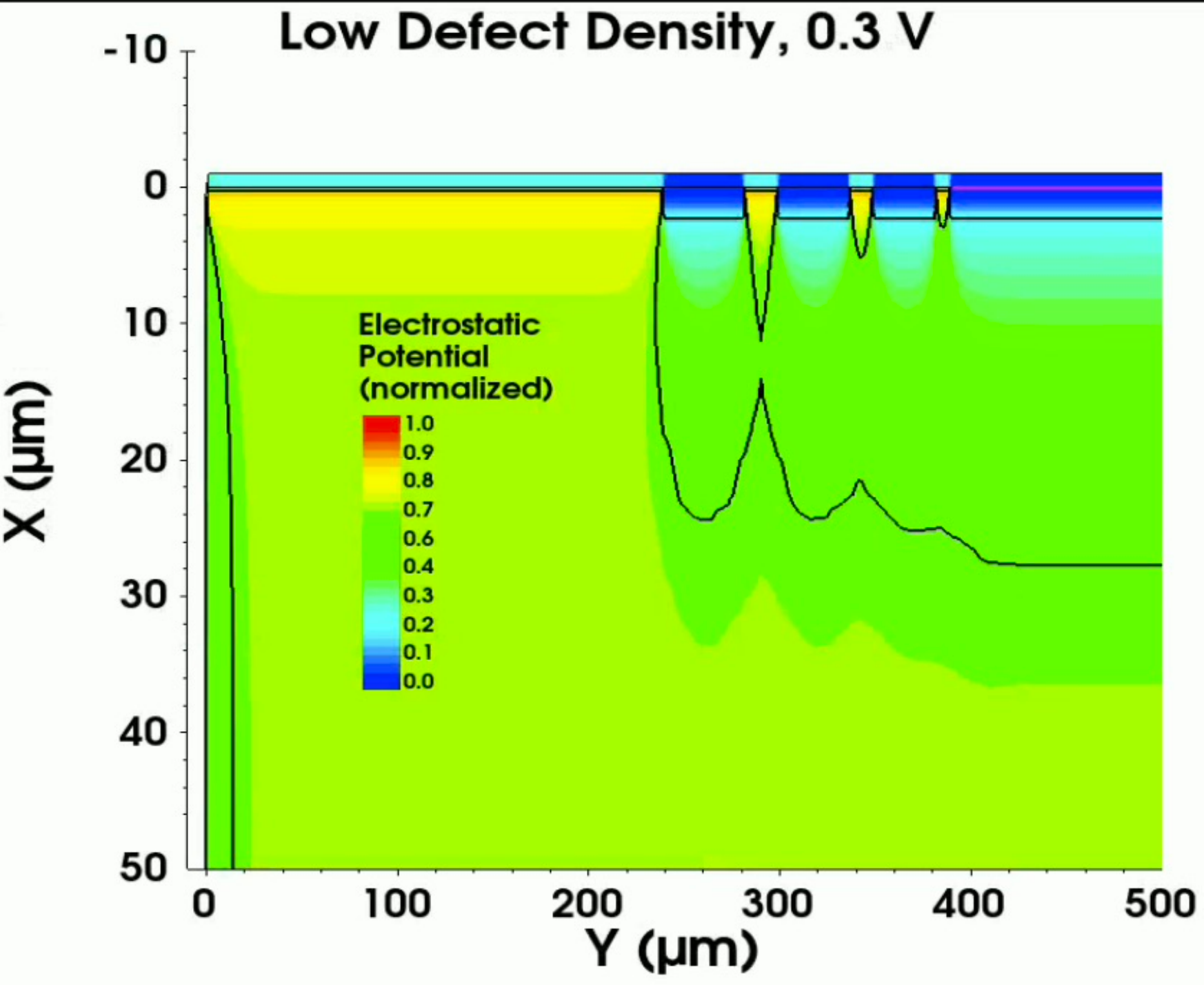}
	\caption{}
	\label{fig:POnNPotential}
\end{subfigure}
\hfill
\begin{subfigure}{.49\textwidth}
        \includegraphics[width=1\textwidth,trim={0 0 0 0.1cm},clip]{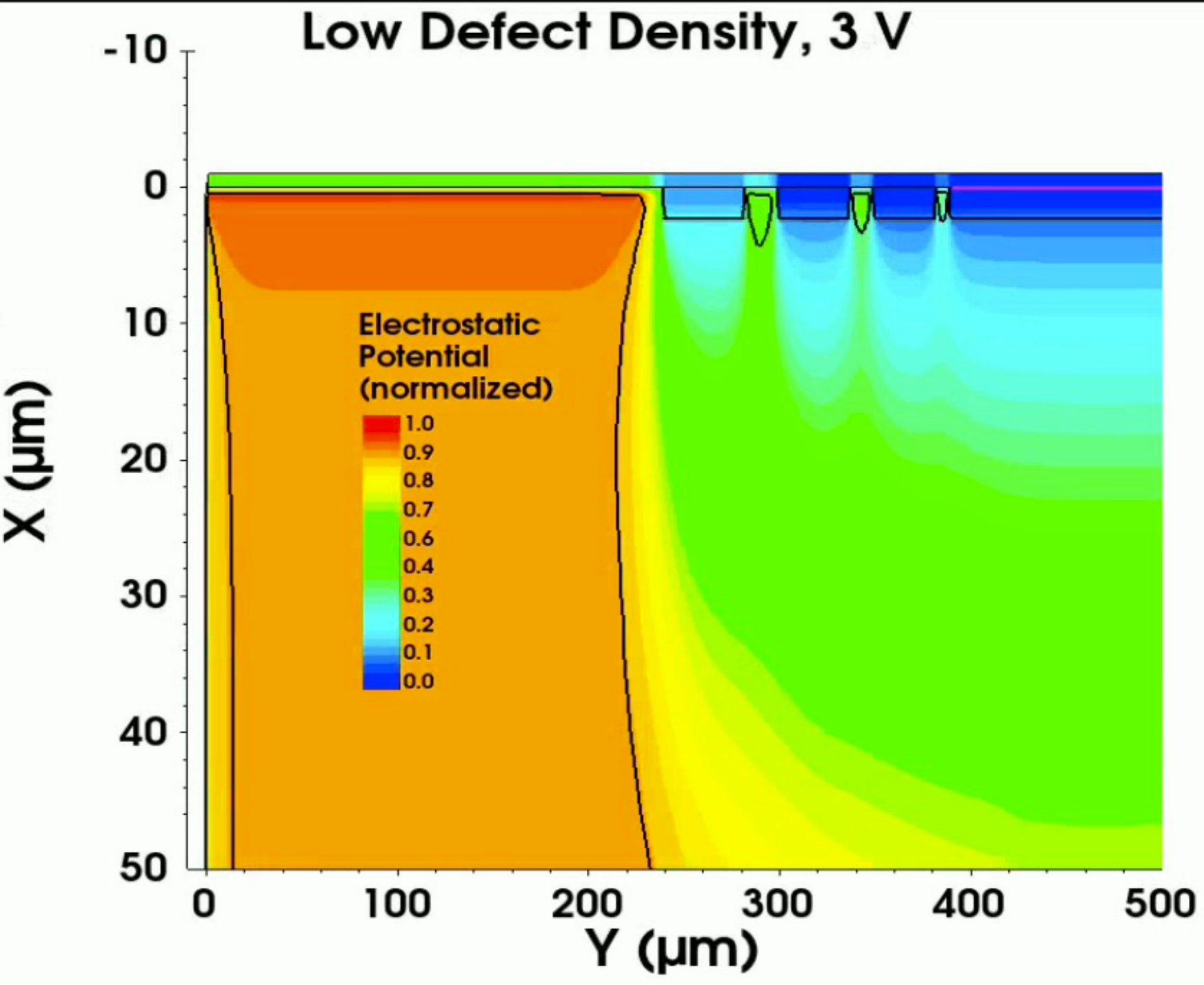}
	\caption{}
	\label{fig:NOnPPotential}
\end{subfigure}
\begin{subfigure}{.49\textwidth}
        \includegraphics[width=1\textwidth,trim={0 0 0 0.1cm},clip]{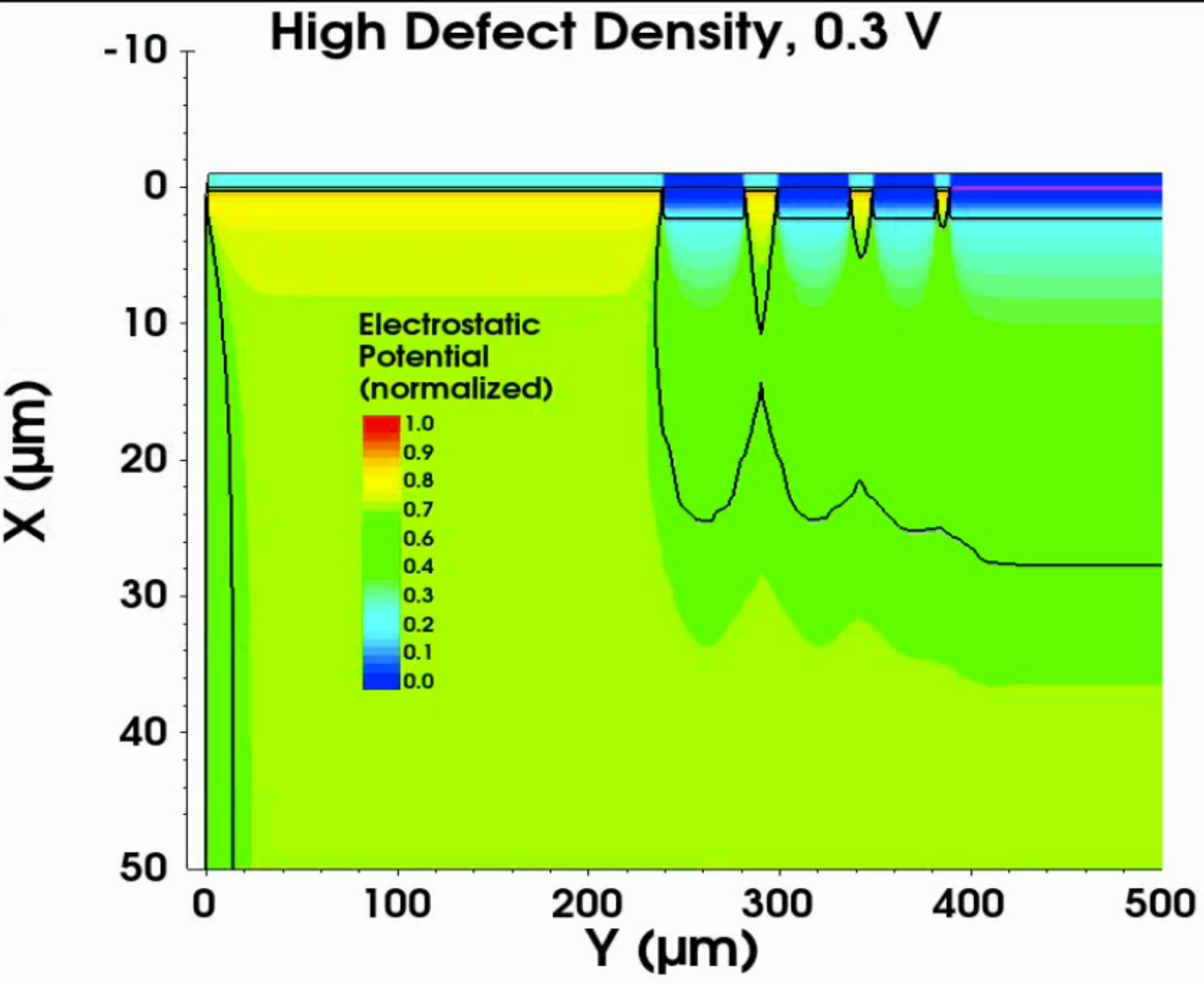}
	\caption{}
	\label{fig:POnNPotential}
\end{subfigure}
\hfill
\begin{subfigure}{.49\textwidth}
        \includegraphics[width=1\textwidth,trim={0 0 0 0.1cm},clip]{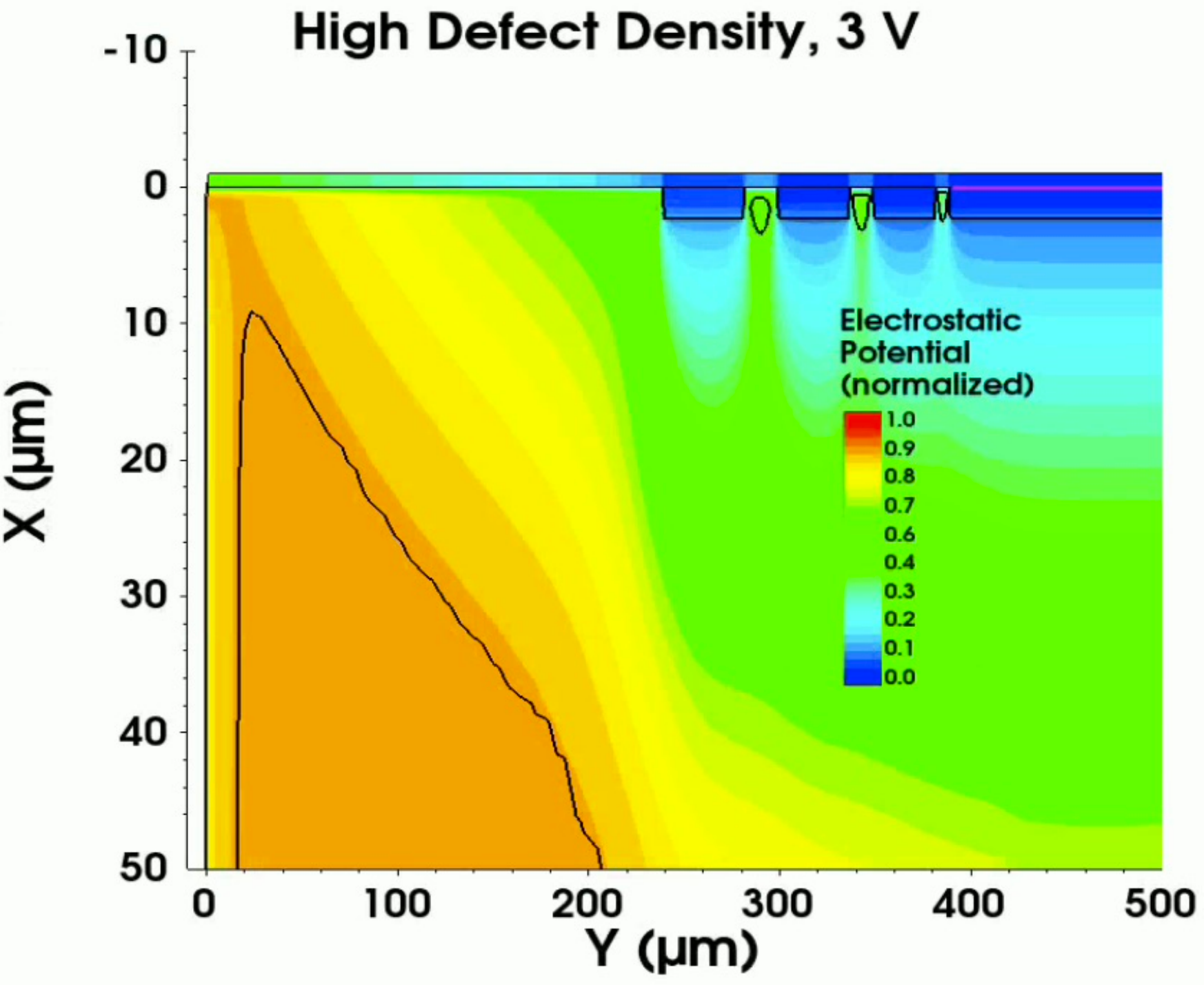}
	\caption{}
	\label{fig:NOnPPotential}
\end{subfigure}
\caption{Normalized electrostatic potential contour plots with depletion region shown with a black contour line with $N_\mathrm{ox} = 2\times 10^{11}~\mathrm{cm^{-2}}$. The upper left corner of an n-on-p device is shown. The 4 different conditions are (a) 0.3 V bias with $N_{\mathrm{don}}=3\times 10^{11}~\mathrm{cm^{-2}}$, (b) 3 V bias with $N_{\mathrm{don}}=3\times 10^{11}~\mathrm{cm^{-2}}$, (c) 0.3 V bias with $N_{\mathrm{don}}=5\times 10^{11}$ and (d) 3 V bias with $N_{\mathrm{don}}=5\times 10^{11}~\mathrm{cm^{-2}}$. The difference between (b) and (d) demonstrates shows that the larger current in (d) is due to edge depletion due to positive interface charge compensating for the p-spray.}
\label{fig:NonPPotentials}
\end{figure}

Figure \ref{fig:NonPPotentials} shows the differing effect on the electrostatic potential and depletion region of different defect densities. The color contours show the electrostatic potential, normalized to the maximum (negative) potential at the device backside. The depletion regions are shown in the black contour lines. Only the top left corner of the device is shown in the plots. At 0.3 V, both low and high defect density devices' electrostatic potentials look nearly the same. But, by 3 V, the edge is depleted for the high density device, and is not for the low density device. In the high density device, the guard rings are all at nearly the same potential. The potential is only slightly reduced between the outer guard ring and the cut edge of the device, whereas for the low density device, it is dropped significantly, and is at nearly the same potential as the backside of the device at the top left corner.

A positive charge density of $5\times 10^{11}~\mathrm{cm^{-2}}$ at the interface is somewhat high, but is not unreasonable to expect, especially for devices fabricated outside of a highly controlled foundry setting. Therefore, it is proposed that the p-spray in these devices did not have a large enough doping concentration to properly isolate the surface, causing the observed behavior. 

\begin{figure}[htbp]
\centering
        \includegraphics[width=.65\textwidth,trim={0 0 0 1.5cm},clip]{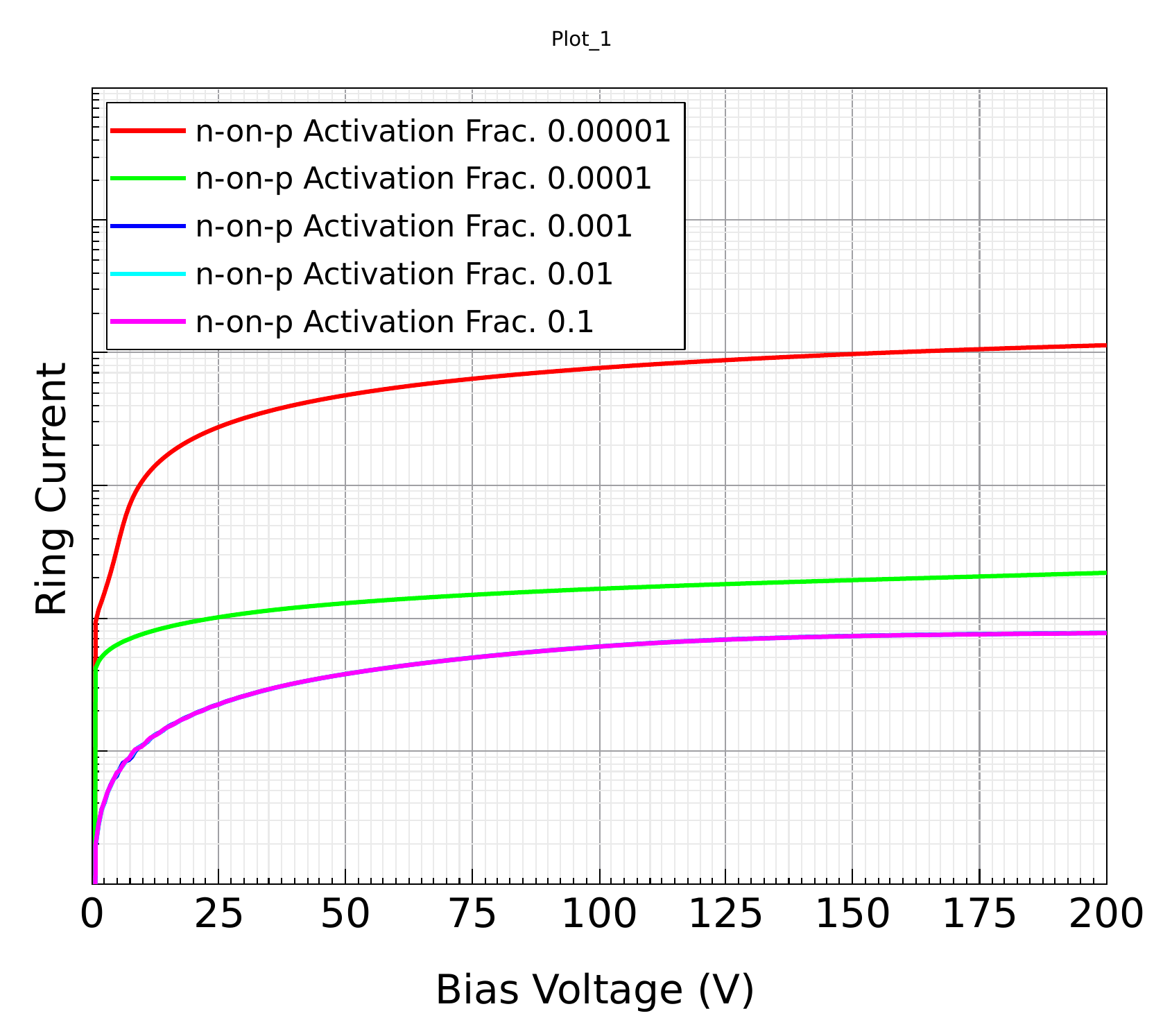}
\caption{Simulated current versus bias voltage curves for $40~\mathrm{\mu m}$ cut distance n-on-p devices with edge implantation and varying fraction of the total implanted dopant activated. Only the two lower activation fraction devices exhibit edge depletion within the 200 V range. The other three curves are nearly identical.}
\label{fig:NOnPSideImplantIV}
\end{figure}

We simulated the post-MWA sidewall implant using a range of activated dopant rates, as described in Section \ref{sec:EdgeDoping}. Most of devices simulated with the p-type sidewall implant exhibited a smaller leakage current with no current increase, consistent with no edge depletion. This is shown in Figure \ref{fig:NOnPSideImplantIV}, which shows the I-V curves for $40~\mathrm{\mu m}$ cut distance devices with activation fractions between $10^{-5}$ and $10^{-1}$. The device with the activation fraction of $10^{-4}$ exhibits a larger current than those with the 3 larger activation fractions, and the device with the activation fraction of $10^{-5}$ has a current comparable to those with no implant at all. Similarly to the p-on-n simulations, even with the lowest activation fractions previously observed for MWA, it is reasonable to expect that the contribution to the leakage current from the edge is removed by the edge implant and MWA.

\section{Outlook}
\label{sec:Outlook}

We have demonstrated the potential for suppressing the edge current using the proposed process on existing prototype sensors with multiple guard rings. The next step to study the proposed technique is the design and fabrication of a dedicated set of active-edge devices using the MWA process. Already, a device with $40~\mathrm{\mu m}$ between the guard ring and diced edge is a significant improvement over typical designs with several hundred microns of extra space. The process could simplify the fabrication of both slim edge devices, with fewer guard rings and very little space between the guard ring and physical edge of the device, and fully edgeless devices with no guard ring, depending on the application.

The implantation process can be further optimized. Devices attached to the holder wafer may have had up to a few microns of the bottom of the edge of the device blocked from implantation by excess photoresist riding up the side of the device when it was pressed onto the holder wafer. Additionally, the MWA power and duration can be optimized, as can the implant parameters.

The n-on-p simulations were able to provide a plausible explanation for the large leakage current seen even at low voltage and with little change based on the edge cut distance, shown in Figure \ref{fig:NonPIV1}, by invoking a relatively large charge due to defects at the $\mathrm{SiO_2-Si}$ interface. However, this is potentially called into question by results showing that a comparable MWA process can lead to significant reduction in interface charge and trap density \cite{Gentry2026}. That result shows that a comparable MWA process leads to a significant reduction in the oxide charge and interface trap density. If that was the case for the devices in this study, it would be expected that after the MWA-only, shown in the I-V curve in Figure \ref{fig:NOnPMWAOnly}, the current for different edge cut distances should shift from looking like the high density defect model in Figure \ref{fig:NOnPIV2} to the low density defect model in Figure \ref{fig:NonPIV1}. However, since the devices roported \cite{Gentry2026} were fabricated by a foundry, whereas the devices in this study were made at SLAC, the kinds of defects in each could be significantly different.

\section{Conclusions}
\label{sec:Conclusions}
Inexpensive and reliable active edge radiation sensors are desirable for a range of applications, including at future high-energy colliders and free-electron laser experiments. Promising results are obtained in this feasibility study for active edge devices fabricated with ion implantation along the cut edge of the device with activation via microwave annealing. Experimental data from test devices demonstrated a significant reduction of the leakage current for both n-on-p and p-on-n sensors, possibly allowing for a tolerable per-pixel current in an edgeless device. Using TCAD simulations qualitative features of the leakage current data before and after the implantation and microwave anneal process are demonstrated and plausible explanations were put forth.

\acknowledgments

This material is based upon work supported by the U.S. Department of Energy, Office of Science, Office of Workforce Development for Teachers and Scientists, Office of Science Graduate Student Research (SCGSR) program. The SCGSR program is administered by the Oak Ridge Institute for Science and Education for the DOE under contract number DE‐SC0014664.

This work was supported in part by the U.S. Department of
Energy under Contract No. DE-AC02-76SF00515 and DE-SC0026381.


\bibliographystyle{JHEP}
\bibliography{main.bib}

@article{Benoit:2009zz,
    author = "Benoit, M. and Lounis, A. and Dinu, N.",
    title = "{Simulation of guard ring influence on the performance of ATLAS pixel detectors for inner layer replacement}",
    doi = "10.1088/1748-0221/4/03/P03025",
    journal = "JINST",
    volume = "4",
    pages = "P03025",
    year = "2009"
}

@inproceedings{Gentry2026,
  author       = {Gentry, Andrew Donald and others},
  title        = {Effects of Microwave Annealing on Surface and Bulk Defects in Silicon Devices},
  booktitle    = {Proceedings of the 21st ``Trento'' Workshop on Advanced Silicon Radiation Detectors},
  year         = {2026},
  address      = {Perugia, Italy},
  month        = feb,
  note         = {Talk presented Feb 19, 2026},
  url          = {https://indico.cern.ch/event/1586892/contributions/6846232/}
}

@article{Batignani:1988as,
title = {Double-sided readout silicon strip detectors for the aleph minivertex},
journal = {Nucl. Instrum. Meth. A},
volume = {277},
number = {1},
pages = {147-153},
year = {1989},
issn = {0168-9002},
doi = {https://doi.org/10.1016/0168-9002(89)90546-9},
url = {https://www.sciencedirect.com/science/article/pii/0168900289905469},
author = {G. Batignani and F. Bosi and L. Bosisio and A. Conti and E. Focardi and F. Forti and M.A. Giorgi and G. Parrini and E. Scarlini and P. Tempesta and G. Tonelli and G. Triggiani}
}

@article{Goessling:1996fn,
    author = "Goessling, C. and others",
    editor = "Holl, P. and Lutz, G. and Richter, Rainer Helmut and Strueder, L. and Longoni, A. and Sampietro, M.",
    title = "{Irradiation tests of double-sided silicon strip detectors optimized for the ATLAS-inner-detector-region}",
    doi = "10.1016/0168-9002(95)01411-X",
    journal = "Nucl. Instrum. Meth. A",
    volume = "377",
    pages = "284--289",
    year = "1996"
}

@INPROCEEDINGS{6154338,
  author={Hansen, T. E. and Ahmed, N. and Ferber, A. and Bouquet, G.},
  booktitle={2011 IEEE Nuclear Science Symposium Conference Record}, 
  title={Edge-on detectors with active edge for X-ray photon counting imaging}, 
  year={2011},
  volume={},
  number={},
  pages={1341-1408},
  doi={10.1109/NSSMIC.2011.6154338}}

@article{Sent-Dev,
    author = {{Synopsis Inc.}},
    title = {Sentaurus Device User Guide},
    year = {2024}
}

@misc{DSG,
    author      = {{DSG Technologies}},
    howpublished = {\url{https://www.dsgtek.com/}}
}

@misc{luxience,
  author       = {{Luxience Technologies}},
  howpublished = {\url{https://www.luxience.com/}},
  note         = {Accessed: 2025-10-01}
}

@article{DallaBetta:2016dqn,
    author = "Dalla Betta, Gian-Franco and others",
    editor = "Bisogni, Maria Giuseppina and Grassi, Marco and Incagli, Marco and Paoletti, Riccardo and Signorelli, Giovanni",
    title = "{The INFN{\textendash}FBK {\textquotedblleft}Phase-2{\textquotedblright} R {\&} D program}",
    eprint = "1612.00626",
    archivePrefix = "arXiv",
    primaryClass = "physics.ins-det",
    doi = "10.1016/j.nima.2015.08.074",
    journal = "Nucl. Instrum. Meth. A",
    volume = "824",
    pages = "388--391",
    year = "2016"
}

@article{Lee2017M2016366,
  title={{Adoption of Hybrid Dicing Technique to Minimize Sawing-Induced Damage during Semiconductor Wafer Separation}},
  author={Seong-Min Lee},
  journal={MATERIALS TRANSACTIONS},
  volume={58},
  number={4},
  pages={530-534},
  year={2017},
  doi={10.2320/matertrans.M2016366}
}

@article{Tsai2022,
    author = {Tsai, Chun-Hsiung and Savant, Chandrashekhar P. and Asadi, Mohammad Javad and Lin, Yu-Ming and Santos, Ivan and Hsu, Yu-Hsiang and Kowalski, Jeffrey and Pelaz, Lourdes and Woon, Wei-Yen and Lee, Chih-Kung and Hwang, James C. M.},
    title = {{Efficient and Stable Activation by Microwave Annealing of Nanosheet Silicon Doped With Phosphorus Above its Solubility Limit}},
    journal = {Applied Physics Letters},
    volume = {121},
    number = {5},
    pages = {052103},
    year = {2022},
    month = {08},
    issn = {0003-6951},
    doi = {10.1063/5.0099083},
    url = {https://doi.org/10.1063/5.0099083}

}

@article{Epix2014,
    author = "Dragone, A. and others",
    title = {{ePix: a class of architectures for second generation LCLS cameras}},
    journal = "Journal of Physics: Conference Series",
    doi = "10.1088/1742-6596/493/1/012012",
    volume = "493",
    pages = "012012",
    year = "2014"}

@article{Epix2016,
    author = "Nishimura, K. and others",
    title = {{Design and Performance of the ePix Camera System}},
    journal = "AIP Conference Proceedings",
    doi = "10.1063/1.4952919",
    volume = "1741",
    year = "2016"}

@article{Fadeyev:2014uaa,
    author = "Fadeyev, V. and others",
    title = "{Update on scribe{\textendash}cleave{\textendash}passivate (SCP) slim edge technology for silicon sensors: Automated processing and radiation resistance}",
    doi = "10.1016/j.nima.2014.05.032",
    journal = "Nucl. Instrum. Meth. A",
    volume = "765",
    pages = "59--63",
    year = "2014"
}

@article{FADEYEV2013260,
title = {Scribe–cleave–passivate (SCP) slim edge technology for silicon sensors},

volume = {731},
pages = {260-265},
year = {2013},
note = {PIXEL 2012},
issn = {0168-9002},
doi = {https://doi.org/10.1016/j.nima.2013.03.046},
url = {https://www.sciencedirect.com/science/article/pii/S0168900213003513},
author = {V. Fadeyev and H.F.-W. Sadrozinski and S. Ely and J.G. Wright and M. Christophersen and B.F. Phlips and G. Pellegrini and S. Grinstein and G.-F. {Dalla Betta} and M. Boscardin and R. Klingenberg and T. Wittig and A. Macchiolo and P. Weigell and D. Creanza and R. Bates and A. Blue and L. Eklund and D. Maneuski and G. Stewart and G. Casse and I. Gorelov and M. Hoeferkamp and J. Metcalfe and S. Seidel and G. Kramberger}

}

@article{Meschini:2016bkd,
    author = "Meschini, M. and Dalla Betta, G. F. and Boscardin, M. and Calderini, G. and Darbo, G. and Giacomini, G. and Messineo, A. and Ronchin, S.",
    title = "{The INFN-FBK pixel R{\&}D program for HL-LHC}",
    doi = "10.1016/j.nima.2016.05.009",
    journal = "Nucl. Instr. Meth. A",
    volume = "831",
    pages = "116--121",
    year = "2016"
}

@article{Ruggiero:2007zzd,
    author = "Ruggiero, G. and Eremin, V. and Noschis, E.",
    editor = "Ambrosi, Giovanni and Bilei, Gian M. and Fano, Livio and Passeri, Daniele and Santocchia, Attilio and Zuccon, Paolo",
    title = "{Planar edgeless silicon detectors for the TOTEM experiment}",
    doi = "10.1016/j.nima.2007.07.110",
    journal = "Nucl. Instrum. Meth. A",
    volume = "582",
    pages = "854--857",
    year = "2007"
}

@article{CMS:2012sda,
    author         = "{CMS Collaboration}",
    editor = "Matzner Dominguez, David Aaron and others",
    collaboration = "CMS",
    title = "{CMS Technical Design Report for the Pixel Detector Upgrade}",
    journal = "CERN-LHCC-2012-016",
    doi = "10.2172/1151650",
    month = "9",
    year = "2012"
}

@article{ATLAS:2017svb,
    author         = "{ATLAS Collaboration}",
    collaboration = "ATLAS",
    title = "{Technical Design Report for the ATLAS Inner Tracker Pixel Detector}",
    journal = "CERN-LHCC-2017-021",
    doi = "10.17181/CERN.FOZZ.ZP3Q",
    year = "2017"
}

@article{Terzo:2021zov,
    author = "Terzo, Stefano and others",
    title = "{Novel 3D Pixel Sensors for the Upgrade of the ATLAS Inner Tracker}",
    doi = "10.3389/fphy.2021.624668",
    journal = "Front. in Phys.",
    volume = "9",
    pages = "2",
    year = "2021"
}

@ARTICLE{Segal:2021,
    
AUTHOR={Segal, Julie  and Kenney, Christopher  and Kowalski, Jeffrey M.  and Kowalski, Jeffrey E.  and Blaj, Gabriel  and Rozario, Lisa  and Hasi, Jasmin  and Dragone, Angelo  and Caragiulo, Pietro  and Rota, Lorenzo },
           
TITLE={Thin-Entrance Window Process for Soft X-Ray Sensors},
          
JOURNAL={Frontiers in Physics},
          
VOLUME={Volume 9 - 2021},
  
YEAR={2021},
  
URL={https://www.frontiersin.org/journals/physics/articles/10.3389/fphy.2021.618390},
  
DOI={10.3389/fphy.2021.618390}
  
}

@ARTICLE{Lu2010,
  author={Lu, Yu-Lun and Hsueh, Fu-Kuo and Huang, Kuo-Ching and Cheng, Tz-Yen and Kowalski, Jeff M. and Kowalski, Jeff E. and Lee, Yao-Jen and Chao, Tien-Sheng and Wu, Ching-Yi},
  journal={IEEE Electron Device Letters}, 
  title={{Nanoscale p-MOS Thin-Film Transistor With TiN Gate Electrode Fabricated by Low-Temperature Microwave Dopant Activation}}, 
  year={2010},
  volume={31},
  number={5},
  pages={437-439},
  doi={10.1109/LED.2010.2042924}}

@article{Segal:2018cyk,
    author = "Segal, Julie D. and Kenney, Christopher J. and Rozario, Lisa and Blaj, Gabriel and Chang, Chu-En and Hasi, Jasmin and Kowalski, Jeffrey M. and Kowalski, Jeffrey E.",
    title = "{Low-temperature Junction Formation for Thinned High Energy Physics Sensors}",
    journal = "{2018 IEEE Nuclear Science Symposium and Medical Imaging Conference}",
    doi = "10.1109/NSSMIC.2018.8824728",
    pages = "8824728",
    year = "2018"
}

@ARTICLE{Lee2009,
  author={Lee, Yao-Jen and Hsueh, Fu-Kuo and Huang, Shih-Chiang and Kowalski, Jeff M. and Kowalski, Jeff E. and Cheng, Alex T. Y. and Koo, Ann and Luo, Guang-Li and Wu, Ching-Yi},
  journal={IEEE Electron Device Letters}, 
  title={{A Low-Temperature Microwave Anneal Process for Boron-Doped Ultrathin Ge Epilayer on Si Substrate}}, 
  year={2009},
  volume={30},
  number={2},
  pages={123-125},
  doi = {10.1109/LED.2008.2009474}
}

@article{ERANEN200985,
title = {{3D Processing on 6in. High Resistive SOI Wafers: Fabrication of Edgeless Strip and Pixel Detectors}},
journal = "Nucl. Instr. Meth. A",
volume = {607},
number = {1},
pages = {85-88},
year = {2009},
note = {Radiation Imaging Detectors 2008},
issn = {0168-9002},
doi = {https://doi.org/10.1016/j.nima.2009.03.243},
url = {https://www.sciencedirect.com/science/article/pii/S016890020900610X},
author = {Simo Eränen and Juha Kalliopuska and Risto Orava and Nick {van Remortel} and Tuula Virolainen}
}

@article{Bates:2013ixa,
    author = "Bates, R. and others",
    title = "{Characterisation of edgeless technologies for pixellated and strip silicon detectors with a micro-focused X-ray beam}",
    doi = "10.1088/1748-0221/8/01/P01018",
    journal = "JINST",
    volume = "8",
    pages = "P01018",
    year = "2013"
}

@article{Parker:1996dx,
    author = "Parker, Sherwood I. and Kenney, Christopher J. and Segal, Julie",
    title = "{3-D: A New architecture for solid state radiation detectors}",
    reportNumber = "UH-511-839-96",
    doi = "10.1016/S0168-9002(97)00694-3",
    journal = "Nucl. Instrum. Meth. A",
    volume = "395",
    pages = "328--343",
    year = "1997"
}

@ARTICLE{Kenney20012405,
	author = {Kenney, Christopher J. and Parker, Sherwood and Walckiers, Edith},
	title = {Results from 3-D silicon sensors with wall electrodes: Near-cell-edge sensitivity measurements as a preview of active-edge sensors},
	year = {2001},
	journal = {IEEE Trans.},
	volume = {48},
	number = {6 III},
	pages = {2405 – 2410},
	doi = {10.1109/23.983250},
	url = {https://www.scopus.com/inward/record.uri?eid=2-s2.0-0035723579&doi=10.1109%2f23.983250&partnerID=40&md5=26cb0a699f39c9eb986914560852458e}
}

@article{KENNEY2006272,
title = {Active-edge planar radiation sensors},
journal = "Nucl. Instrum. Meth. A",
author = {C.J. Kenney and J.D. Segal and E. Westbrook and Sherwood Parker and J. Hasi and C. {Da Vià} and S. Watts and J. Morse},
volume = {565},
number = {1},
pages = {272-277},
year = {2006},
note = {Proc. of the Int. Workshop on Semiconductor Pixel Detectors for Particles and Imaging},
issn = {0168-9002},
doi = {https://doi.org/10.1016/j.nima.2006.05.012},
url = {https://www.sciencedirect.com/science/article/pii/S0168900206007765}
}

@article{BOMBEN201341,
title = {Development of edgeless n-on-p planar pixel sensors for future {ATLAS} upgrades},
author = {Marco Bomben and Alvise Bagolini and Maurizio Boscardin and Luciano Bosisio and Giovanni Calderini and Jacques Chauveau and Gabriele Giacomini and Alessandro {La Rosa} and Giovanni Marchiori and Nicola Zorzi},
journal = "Nucl. Instrum. Meth. A",
volume = {712},
pages = {41-47},
year = {2013},
issn = {0168-9002},
doi = {https://doi.org/10.1016/j.nima.2013.02.010},
url = {https://www.sciencedirect.com/science/article/pii/S0168900213001897}
}

@article{KOYBASI2020163176,
title = {{Edgeless Silicon Sensors Fabricated Without Support Wafer}},
journal = "Nucl. Instrum. Meth. A",
volume = {953},
pages = {163176},
year = {2020},
issn = {0168-9002},
doi = {https://doi.org/10.1016/j.nima.2019.163176},
url = {https://www.sciencedirect.com/science/article/pii/S0168900219314792},
author = {Ozhan Koybasi and Jiaguo Zhang and Angela Kok and Anand Summanwar and Marco Povoli and Lars Breivik and Anna Bergamaschi and Bernd Schmitt}
}

\end{document}